\documentclass{rspublic}

\usepackage{indentfirst}

\usepackage[english]{babel}
\usepackage[latin1]{inputenc}

\usepackage[dvips,final]{graphicx}
\usepackage{subfigure}

\usepackage[sort,sectionbib]{natbib}

\usepackage{xspace} 
\usepackage{amsmath,amssymb,amsthm}

\newcommand{\pd}[2]{\frac{\partial#1}{\partial#2}}

\newcommand{\abs}[1]{\left|#1\right|}
\newcommand{\eps}{\varepsilon}
\newcommand{\m}{\mu^2}
\newcommand{\mm}{\mu^4}

\def\R{\mathbb{R}}
\def\u{\mathbf{u}}
\def\k{\mathbf{k}}
\def\x{\mathbf{x}}

\def\g{\mathbf{g}}

\def\Id{\mathrm{Id}}

\begin{document}

\title{Energy of tsunami waves generated by bottom motion}

\author{Denys Dutykh\footnote{now at LAMA, University of Savoie,
CNRS, Campus Scientifique, 73376 Le Bourget-du-Lac Cedex, France}, Fr\'ed\'eric Dias}

\affiliation{CMLA, ENS Cachan, CNRS, PRES UniverSud,
61, avenue du President Wilson, 94230 Cachan cedex, France}

\label{firstpage}

\maketitle

\begin{abstract}{water waves, shallow-water equations, tsunami energy}

In the vast literature on tsunami research, few articles have been devoted to energy issues. A theoretical investigation on the energy of waves generated by bottom motion is performed here. We start with the full incompressible Euler equations in the presence of a free surface and derive both dispersive and non-dispersive shallow-water equations with
an energy equation. It is shown that dispersive effects only appear at higher order in the energy budget. Then we solve
the Cauchy--Poisson problem of tsunami generation for the linearized water
wave equations. Exchanges between potential and kinetic energies are clearly revealed.
\end{abstract}

\section{Introduction}

Oceanic waves can be devastating as shown by recent events. Whilst some areas are more vulnerable than others, the recent history shows that catastrophic waves can hit even where they are not expected. The tsunami waves generated by the huge undersea earthquake in Indonesia on 26 December 2004 caused devastation across most of the coasts of the Bay of Bengal. The tsunami waves generated by the massive submarine landslide in Papua-New Guinea on 17 July 1998 as well as the 17 July 2006 Java tsunami and the 2 April 2007 Solomon Islands tsunami also caused devastation, but on a smaller scale.  Unfortunately, such cataclysmic tsunamis are likely to be generated again by earthquakes, massive landslides or volcano eruptions \citep{Syno2006}.

Information on tsunami energy can be obtained by applying the normal mode representation of tsunami waves, as introduced by 
\cite{Ward80}. For example, \cite{Okal03} considers the total energy released into tsunami waves. He obtains expressions for the energy of tsunamis (see his expressions (31) for a tsunami generated by an earthquake and (36) for a tsunami generated by a landslide). In the case of a landslide, he computes the ratio between tsunami energy and total change in energy due to the slide. In the present paper, we use the incompressible fluid dynamics equations. Tsunamis have traditionally been considered as non-dispersive long waves. However various types of data (bottom pressure records \citep{Ritsema1995};
satellite data \citep{Kulikov2005}; hydrophone records \citep{OTR2007}) indicate that tsunamis are made up of a very long dispersive wave train, especially when they have enough time to propagate. These waves travel across the ocean surface in all directions away from the generation region. Recent numerical computations using dispersive wave models such as the Boussinesq equations show as much as 20\% reduction of tsunami amplitude in certain locations due to dispersion \citep{Dao2007}. But one has to be careful with the interpretation of satellite data: as indicated by \cite{KS2006}, the mid-ocean steepness of the 2004 Sumatra tsunami measured from satellite altimeter data was less than 10$^{-5}$. Nonlinear dispersive theory is necessary only when examining steep gravity waves, which is not the case in deep water. 

The wavelength of tsunamis and, consequently, their period depend essentially on the source mechanism. If the tsunami is generated by a large and shallow earthquake, its initial wavelength and period will be greater. On the other  hand, if the tsunami is caused by a landslide (which happens less commonly but can be devastating as well), both its initial wavelength and period will be shorter as indicated for example by \cite{Kulikov1996}. From these empirical considerations one can conclude that dispersive effects are a priori more important for submarine landslide and slump scenarios than for tsunamigenic earthquakes.

Once a tsunami has been generated, its energy is distributed throughout the water column. Clearly, the more water is displaced,
the more energetic is the tsunami (compare for example the December 2004 and March 2005 Sumatra tsunamis). Due to the large scale of this natural phenomenon and limited power of computers, tsunami wave modellers have to adopt simplified models which reduce a fully three-dimensional (3D) problem to a two-dimensional (2D) one. This approach is natural, since in the case of very long waves the water column moves as a whole. Consequently the flow is almost 2D. Among these models one can mention the nonlinear shallow water equations (SWE), Boussinesq type models, the Green-Naghdi and Serre equations. There is a wide variety of models, depending on whether or not the effects of run-up/run-down, bottom friction, turbulence, Coriolis effects, tidal effects, etc,
are included.

Today scientists can easily predict when a tsunami will arrive at various places by knowing source characteristics and bathymetry data along the paths to those places. Unfortunately one does not know as much about the energy propagation of such waves. Obviously tsunami amplitude is enhanced over the major oceanic ridges. \cite{Titov} clearly describe the waveguide type
effect from mid-ocean ridges that has funnelled the 2004 megatsunami away from the tip of Africa. As emphasized by \cite{Kowalik2007}, travel-time computation based on the first arrival time may lead to errors in the prediction of tsunami arrival time as higher energy waves propagate slower along ridges. At the beginning, the energy is essentially potential, although it depends on the generation mechanism. Then it redistributes itself into
half kinetic and half potential energies. Finally, it converts its potential component into kinetic energy. How do these conversions take place? The purpose of this study is to shed some light on this topic and to see if the importance of dispersion in tsunamis can be studied by looking at the energy rather than at wave profiles.

Previous researchers have considered the topic of tsunami wave energy. 
\cite{Kajiura1970,Dotsenko1997,Velichko2002} studied the energy exchange between the solid bottom and the overlying water associated with the bottom deformation. There were recent attempts to obtain equations for tsunami energy propagation. We can mention here the work of \cite{Tinti2000} devoted to idealized theoretical cases and the work of \cite{Kowalik2007} using the energy flux point of view to study the changes in the 2004 tsunami signal as it travelled from Indonesia to the Pacific Ocean. We believe that these models can be improved, given the present state of the art in wave modelling.

A point of interest is that some of the equations used for wave modelling have an infinite number of conserved quantities. There has been some confusion in the literature on which quantities can be called energy. Indeed there is here an interesting question. In incompressible fluid mechanics, the internal energy equation is decoupled from the equation of continuity and from the fundamental law of dynamics. It is used only when one is interested in computing the temperature field once the velocity distribution is known. In addition to the internal energy equation, one can write a total energy (internal energy + kinetic energy) equation, or a total enthalpy equation. The confusing part is that for perfect fluids one usually defines the total energy differently: it is the sum of internal energy, kinetic energy, and potential energies associated to body forces such as gravitational forces and to the pressure field. If in addition the fluid is incompressible, then the internal energy remains constant. In the classical textbooks on water waves \citep{Stoker1958, Johnson1997}, one usually introduces the energy $E$ as the sum of kinetic and potential energies and then looks for a partial differential equation giving the time derivative $dE/dt$ (incidentally the meaning of $d/dt$ is not always clearly defined). In any case, when one uses a depth-integrated model such as the nonlinear SWE, one can compute the energy a posteriori (the potential energy is based on the free-surface elevation and the kinetic energy on the horizontal velocity). But one can also apply the nonlinear shallow water assumptions to the full energy equation to begin with. Then one obtains a nonlinear shallow water approximation of the energy equation. Are these two approaches equivalent? We show that the answer is no.

First we present the energy equation in three different forms: full water wave equations, dispersive SWE and non-dispersive SWE. Surprisingly, the energy equation is the same for dispersive
and non-dispersive SWE at leading order. Then we present some numerical computations over a flat bottom. It allows us to concentrate on the generation process and the energy transfer through the moving seabed. We refer to \cite{VOLNA} for simulations of some real world events
including energy pumping and its transformation over uneven bathymetry, using the SWE with energy. Finally we present the energy
equation in a fourth form: the linearized dispersive water wave equations. We solve the
Cauchy--Poisson problem of tsunami generation. Exchanges between potential and kinetic energies are clearly revealed.



\section{Derivation of the energy equation}

\begin{figure}
	\centering
		\includegraphics[width=0.98\textwidth]{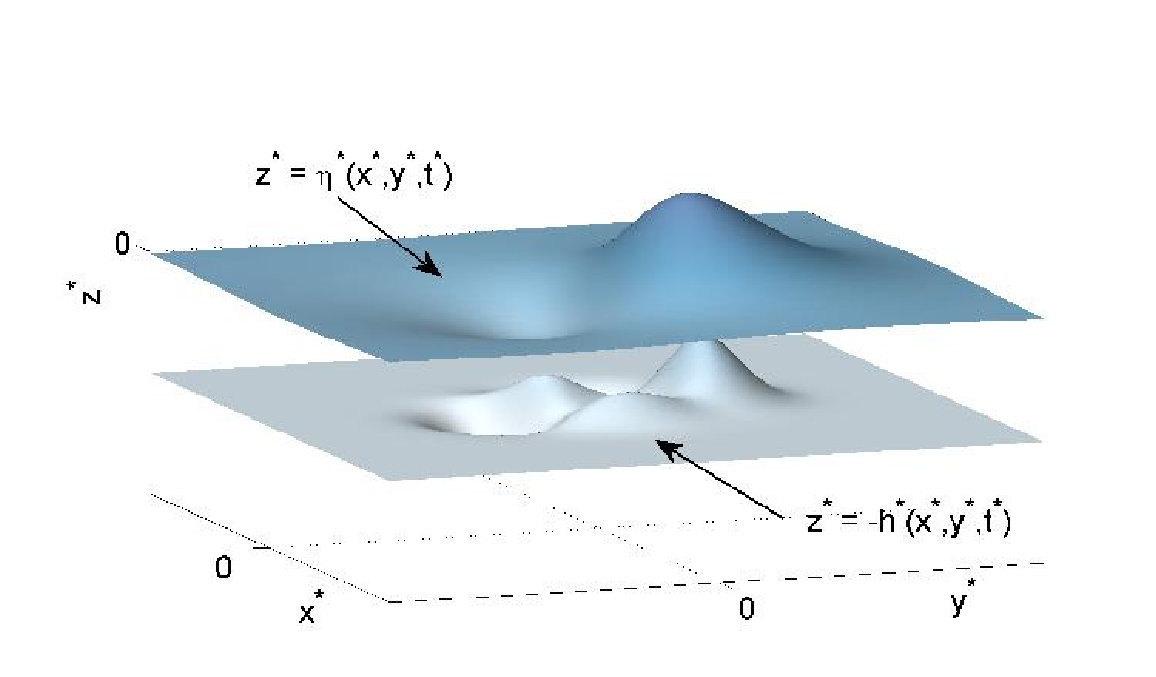}
	\caption{Sketch of the 3D fluid domain for wave generation by a moving bottom.}
	\label{fig:fluiddomain}
\end{figure}

Consider the 3D fluid domain shown in 
Figure \ref{fig:fluiddomain}. It is bounded above
by the free surface $z^*=\eta^*(x^*,y^*,t^*)$ and below by the solid boundary with prescribed motion
$z^*=-h^*(x^*,y^*,t^*)$. A Cartesian coordinate system with the $z^*-$axis pointing vertically upwards and the $x^*Oy^*-$plane 
coinciding with the still-water level is chosen.

The fluid is assumed to be inviscid. Its motion is governed by the 3D Euler 
equations, written here in their incompressible form (see for example \cite{Gisler2008} for the compressible
counterpart, after replacing $\rho$ by $p$ in the pressure term of his Eq. 3):
\begin{eqnarray}\label{eq:EulerIncompr1}
  \nabla\cdot\u^* & = & 0, \\
  \pd{\u^*}{t^*} + \nabla\cdot\left(\u^*\otimes\u^* + \frac{p^*}{\rho^*}\Id\right) & = & \g,\label{eq:EulerIncompr2} \\
  \pd{e^*}{t^*} + \nabla\cdot\left[\left(e^*+\frac{p^*}{\rho^*}\right)\u^*\right] & = & 0,\label{eq:EulerIncompr3}
\end{eqnarray}
where $\rho^*$ is the fluid density, $\u^*=(u^*,v^*,w^*)$ the velocity vector, $e^*$ the sum of kinetic energy density $e_K^*=\frac12\left|\u^*\right|^2$ and potential energy density $e_P^*=gz^*$,
$p^*$ the pressure and $\g$ the acceleration due to gravity. In the present study
$\g=(0,0,-g)$. 
For incompressible flows, the energy equation (\ref{eq:EulerIncompr3}) is redundant. Indeed it can be obtained from Eq. (\ref{eq:EulerIncompr2}). However we keep it since it 
is not equivalent to derive shallow water equations with or without the energy equation. 

Equations (\ref{eq:EulerIncompr1})--(\ref{eq:EulerIncompr3}) have to be completed 
by the kinematic and dynamic boundary conditions. Since surface tension effects are not important for long 
waves, the dynamic boundary condition on the free surface reads
\begin{equation}\label{eq:dynbound}
  p^* = p^*_s, \quad \mbox{at} \;\; z^* = \eta^*.
\end{equation}
Later we will replace the surface pressure $p^*_s$ by 0 but we keep it arbitrary for now. 

The kinematic boundary conditions on the free surface and at the seabed are, respectively,
\begin{equation}\label{eq:kinfreesurf}
  w^* = \pd{\eta^*}{t^*} + u^*\pd{\eta^*}{x^*} + v^*\pd{\eta^*}{y^*}, \qquad z^* = \eta^*,
\end{equation}
\begin{equation}\label{eq:kinbottom}
  w^* = -\pd{h^*}{t^*} - u^*\pd{h^*}{x^*} - v^*\pd{h^*}{y^*}, \qquad z^* = -h^*.
\end{equation}
Below we denote the horizontal gradient by $\nabla_{\perp}$ and the horizontal velocity by $\u^*_{\perp}$.
After a few manipulations and integration across the water column from bottom to top, one can write the following global energy equation:
\begin{equation}\label{energyequation}
\pd{E^*}{t^*} + \nabla_{\perp} \cdot \Phi^* + P^* = 0,
\end{equation}
where $E^*$ is the sum of kinetic and potential energies in the flow, per unit horizontal area, $\Phi^*$ the horizontal energy flux vector, and $P^*$ the net energy input due to the pressure forces doing work on the upper and lower boundaries of the fluid. They are given by the following expressions:
\begin{eqnarray}\label{realenergy}
E^* & = & \int_{-h^*}^{\eta^*} \left( \frac12 \rho^* \left|\u^*\right|^2 + \rho^* g z^* \right) dz^*, \\
\Phi^* & = & \int_{-h^*}^{\eta^*} \u^*_{\perp} \left( \frac12 \rho^* \left|\u^*\right|^2 + p^* + \rho^* g z^* \right) dz^*, \\
P^* & = & p^*_s \eta^*_{t^*} + p^*_b h^*_{t^*},
\end{eqnarray}
where $p^*_s$ is the pressure exerted on the free surface and $p^*_b$ the bottom pressure. 
In the case of a stationary bottom boundary and of a free surface on which the pressure vanishes, then as expected the net energy input $P^*$ is identically zero. Energy can be brought to the system by a moving bottom or by a pressure disturbance on the free surface. From now on, we take $p^*_s=0$.

\subsection{Dimensionless equations}

The problem of tsunami propagation possesses two characteristic length scales: 
the average water depth $h_0$ for the vertical dimension and a typical wavelength $l$ for the horizontal dimensions.
It is classical to introduce the following dimensionless variables. The scaling for the independent variables is
\begin{equation*}
 x = \frac{x^*}{l}, \quad 
 y = \frac{y^*}{l}, \quad
 z = \frac{z^*}{h_0}, \quad
 t = \frac{\sqrt{gh_0}}{l}t^*.
\end{equation*}
In order to introduce the dimensionless dependent variables we need one more parameter, the typical wave amplitude $a$:
\begin{equation*}
  u = \frac{h_0}{a\sqrt{gh_0}}u^*, \quad
  v = \frac{h_0}{a\sqrt{gh_0}}v^*, \quad
  w = \frac{h_0}{l}\frac{h_0}{a\sqrt{gh_0}}w^*, \quad
  \eta = \frac{\eta^*}{a}, \quad
  h = \frac{h^*}{h_0},
\end{equation*}
\begin{equation*}
  \pi = \frac{p^*+\rho^* gz^*}{\rho^* g a}, \quad p = \frac{p^*}{\rho^* g h_0}, \quad
  e = \frac{e^*}{g h_0}.
\end{equation*}
The hydrostatic pressure $-\rho^* gz^*$ has been incorporated into $\pi$.

The following dimensionless parameters, which are assumed to be small, are introduced:
\begin{equation*}
  \eps := {a}/{h_0}, \quad \mu := {h_0}/{l}.
\end{equation*}
The parameter $\eps$ represents the relative importance of nonlinear terms and $\mu$ measures
dispersive effects. Note that 
\begin{equation*}
e = \frac{1}{2} \eps^2 (u^2+v^2) + \frac{1}{2} \frac{\eps^2}{\mu^2} w^2 + z.
\end{equation*}

The Euler equations of motion (\ref{eq:EulerIncompr1})--(\ref{eq:EulerIncompr3}) become in dimensionless form
\begin{equation}\label{eq:continuity}
  \m(u_x + v_y) + w_z = 0,
\end{equation}
\begin{equation}\label{eq:xmomentum}
  \m u_t + \eps\m(u^2)_x + \eps\m(uv)_y + \eps(uw)_z + \m \pi_x = 0,
\end{equation}
\begin{equation}\label{eq:ymomentum}
  \m v_t + \eps\m(uv)_x + \eps\m(v^2)_y + \eps(vw)_z + \m \pi_y = 0,
\end{equation}
\begin{equation}\label{eq:zmomentum}
  \m w_t + \eps\m(uw)_x + \eps\m(vw)_y + {\eps}(w^2)_z + \m \pi_z = 0,
\end{equation}
\begin{equation}\label{eq:energy}
  \m e_t + \eps\m \left((e+p)u\right)_x + \eps\m \left((e+p)v\right)_y + \eps\left((e+p)w\right)_z = 0.
\end{equation}

In dimensionless form the boundary conditions (\ref{eq:dynbound})--(\ref{eq:kinbottom}) become
\begin{equation}\label{eq:dynnondim}
  \pi = \eta, \qquad z = \eps \eta,
\end{equation}
\begin{equation}\label{eq:kinfreesurfnondim}
  w = \m\eta_t + \eps\m u\eta_x + \eps\m v\eta_y, \qquad z = \eps\eta,
\end{equation}
\begin{equation}\label{eq:kinbottomnondim}
  \eps w = -\m h_t - \eps\m u h_x - \eps\m v h_y, \qquad z = -h.
\end{equation}

In the case of a static bottom $h = h(x,y)$, the order of magnitude of the
vertical velocity $w$ at the bottom is $O(\m)$. With a moving bathymetry the behaviour is different:
$$
  \left. w\right|_{z=-h} = -\frac{\m}{\eps} h_t + O(\m).
$$


\subsection{Integration over the depth}

Next we reduce the above 3D problem (\ref{eq:continuity})--(\ref{eq:kinbottomnondim}) into a 2D 
one by integrating all the equations over the water 
column. We begin with the continuity equation (\ref{eq:continuity}) which we integrate with respect 
to $z$ from $-h$ to $\eps\eta$. Taking into account the boundary conditions 
(\ref{eq:kinfreesurfnondim}) and (\ref{eq:kinbottomnondim}), one obtains
\begin{equation}\label{eq:contint}
  \eta_t + \pd{}{x}\int\limits_{-h}^{\eps\eta} u\;dz +
  \pd{}{y}\int\limits_{-h}^{\eps\eta} v\;dz = - \frac1{\eps} h_t.
\end{equation}

A source term appears in equation (\ref{eq:contint}) due to the moving
bathymetry. For tsunami generation, the function $h^*$ can be represented as follows:
$$
  h^*(x^*,y^*,t^*) = h^*_0(x^*,y^*) - \zeta^* (x^*,y^*,t^*),
$$
where $h^*_0(x^*,y^*)$ is the static sea bed profile and $\zeta^*(x^*,y^*,t^*)$ the bottom displacement due for example to 
coseismic displacements or landslides. In nondimensional form this representation
takes the form:
$$
  h(x,y,t) = h_0(x,y) - \eps \zeta(x,y,t),
$$
since the bottom variation must be of the same order of magnitude
as the typical wave amplitude $a$: $\zeta^*(x^*,y^*,t^*)=a\zeta(x,y,t)$. Differentiating with respect to time yields
\begin{equation*}
  \frac1{\eps} \pd{h}{t} = -\pd{\zeta}{t} = O(1).
\end{equation*}
Below we will replace $\frac1{\eps} h_t$ by $-\zeta_t$.


Integrating the vertical momentum conservation equation (\ref{eq:zmomentum}) yields
\begin{equation}\label{eq:pressbot}
  \left.\pi\right|_{z=-h} = \eta + 
  \pd{}{t}\int\limits_{-h}^{\eps\eta} w\;dz + 
  \eps\pd{}{x}\int\limits_{-h}^{\eps\eta} u w\;dz + 
  \eps\pd{}{y}\int\limits_{-h}^{\eps\eta} v w\;dz.
\end{equation}

A more general expression for the pressure can be obtained if we integrate equation 
(\ref{eq:zmomentum}) from $z$ to $\eps\eta$ and use the boundary conditions 
(\ref{eq:dynnondim}) and (\ref{eq:kinfreesurfnondim}):
\begin{equation}\label{eq:pressatz}
  \pi = \eta + 
  \pd{}{t}\int\limits_{z}^{\eps\eta} w\;dz + 
  \eps\pd{}{x}\int\limits_{z}^{\eps\eta} u w\;dz + 
  \eps\pd{}{y}\int\limits_{z}^{\eps\eta} v w\;dz - \frac{\eps}{\m} w^2.
\end{equation}

The vertical velocity $w$ is obtained by integrating the continuity equation (\ref{eq:continuity})
from $-h$ to $z$ and applying the seabed kinematic condition (\ref{eq:kinbottomnondim}):
\begin{equation}\label{eq:verticalcelerityint}
  w = \m \zeta_t - \m\left(
  \pd{}{x}\int\limits_{-h}^z u\; dz + 
  \pd{}{y}\int\limits_{-h}^z v\; dz \right).
\end{equation}

Finally, the integration of the equation of energy conservation (\ref{eq:energy}) yields
\begin{equation}\label{eq:energyint}
  \pd{}{t}\int\limits_{-h}^{\eps\eta} e\; dz +
  \eps\pd{}{x}\int\limits_{-h}^{\eps\eta}(e+p)u\;dz + 
  \eps\pd{}{y}\int\limits_{-h}^{\eps\eta}(e+p)v\;dz - \eps
  \left.p\right|_{z=-h} \zeta_t = 0,
\end{equation}
which is nothing else than the dimensionless counterpart of Eq. (\ref{energyequation}).

All the equations derived above are exact and no
assumption has been made about the orders of magnitude of $\eps$ and $\mu$.

\subsection{The nonlinear shallow-water equations with energy equation}

In Appendix A, we briefly summarize the derivation of various systems of shallow-water wave equations. The non-dispersive SWE are obtained by taking the limit $\mu \to 0$. Here we provide the dispersive and non-dispersive 
SWE in their conservative forms \citep{Eskilsson} with dimensions, based on the
depth-averaged horizontal velocity $\bar\u^*$. The total water depth $h^*+\eta^*$ is denoted by $H^*(x^*,y^*,t^*)$.
The definition of $E^*$ is given by Eq. (\ref{realenergy}).

\subsubsection{SWE with dispersion and energy equation}
\begin{eqnarray}\label{eq:dispEskilsson}
\pd{H^*}{t^*} + \nabla \cdot (H^*\bar\u^*) & = & 0, \\
\pd{(H^*\bar\u^*)}{t^*} + \nabla \cdot \left(H^* \bar\u^* \otimes \bar\u^* + \frac{1}{2}gH^{*2}\right) + \hspace{1cm} & & \nonumber \\
\left( \frac{h^{*3}}{6}\nabla\left(\nabla\cdot\left(\frac{H^*\bar\u^*}{h^*}\right)\right) - \frac{h^{*2}}{2}\nabla\left(\nabla\cdot\left(H^*\bar\u^*\right)\right) \right)_{t^*} & = & gH^* \nabla h^*, \\
\pd{E^*}{t^*} + \nabla \cdot \left( \bar\u^*\left(E^* + \frac{1}{2}\rho^* gH^{*2}\right)\right) & = & 
\rho^* g H^* \pd{\zeta^*}{t^*}.
\end{eqnarray}

\subsubsection{SWE without dispersion and with energy equation}
\begin{eqnarray}\label{eq:nondispEskilsson1}
\pd{H^*}{t^*} + \nabla \cdot (H^*\bar\u^*) & = & 0, \\
\pd{(H^*\bar\u^*)}{t^*} + \nabla \cdot \left(H^* \bar\u^* \otimes \bar\u^* + \frac{1}{2}gH^{*2}\right) & = & gH^* \nabla h^*, 
\label{eq:nondispEskilsson2} \\
\pd{E^*}{t^*} + \nabla \cdot \left( \bar\u^*\left(E^* + \frac{1}{2}\rho^* gH^{*2}\right)\right) & = & 
\rho^* g H^* \pd{\zeta^*}{t^*}. \label{eq:nondispEskilsson3}
\end{eqnarray}
Between (i) and (ii), only the equation for the evolution of the horizontal velocity (\ref{eq:nondispEskilsson2}) differs. In particular, it is interesting to note (this is the first main result of the present paper) that the energy
equations are the same in the dispersive and non-dispersive cases. Differences appear only at higher order (terms of order $O(\eps\m)$). The physical meaning is that the vertical velocity as well as the non-uniform structure of the horizontal
velocity appear only at the next order. An additional remark can be made about the hyperbolic structure of the system 
(\ref{eq:nondispEskilsson1})--(\ref{eq:nondispEskilsson3}). We restrict our observations to one space dimension. Let us assume that a shock wave propagates from left to right at velocity $s > 0$. The states before and after the discontinuity are
denoted by $(H^*_{l,r}, \bar u^*_{l,r}, E^*_{l,r})$ correspondingly. From entropy considerations one can conclude that for an admissible state $H^*_l > H^*_r$. For a general system of conservation laws $v_t + \partial_x f(v) = 0$, the Rankine-Hugoniot relations have the simple form $f(v_r) - f(v_l) = s (v_r - v_l)$.

Some simple algebraic calculations yield the following relations:
\begin{equation*}
  s = \frac{H^*_l \bar u^*_l - H^*_r \bar u^*_r}{H^*_l - H^*_r}, \quad
  \bar u^*_r = \bar u^*_l \pm \sqrt{\frac{1}{2}g(H_l^{*2} - H_r^{*2})\Bigl(\frac{1}{H^*_r} - \frac{1}{H^*_l}\Bigr)},
\end{equation*}
\begin{equation}\label{eq:energyshock}
  \frac{E^*_r}{H_r^*} = \frac{E^*_l}{H_l^*} - \frac{1}{2}\rho^* g\Bigl(\frac{1}{H^*_r} - \frac{1}{H^*_l}\Bigr)
  \frac{H_r^{*2}\bar u^*_r - H_l^{*2}\bar u^*_l}{\bar u^*_r - \bar u^*_l}.
\end{equation}
These formulas relate left and right states connected in the $(H^*, \bar u^*, E^*)$ space by a shock wave. The first two relations are well known. What is new to our knowledge is the formula (\ref{eq:energyshock}) which gives an insight into the energy states in a shock wave. In practice, they can be used for the theoretical analysis of bores and as a validation test for nonlinear SWE codes (with energy equation). 

\section[Simulations of energy]
{Simulations of energy}

Next we illustrate the main features of energy evolution in tsunami generation. The importance of dispersive effects
strongly depends on the extent of the source area (the smaller the source the stronger the dispersive effects) and
the ocean depth in the source area \citep{Kervella2007}. We restrict
our study to the non-dispersive SWE (\ref{eq:nondispEskilsson1})--(\ref{eq:nondispEskilsson3}).
We solve these equations numerically with a finite volume method \citep{VOLNA}.

While the common practice in modeling tsunami generation consists in translating the 
initial sea bottom deformation to the water surface, thus neglecting all dynamical 
effects, we prefer to include some dynamics in the process in an effort to be closer to
what happens in reality \citep{Dutykh2006a}. We construct the bottom motion by multiplying
Okada's static solution $\zeta_{OK}^*(x^*,y^*)$\footnote{Okada's solution is a steady analytical solution for the seafloor displacement
following an underwater earthquake, based on dislocation theory in an elastic half-space \citep{Okada85}.} by a function of time \citep{Hammack1973}: 
\begin{equation*}
  h^*(x^*,y^*,t^*) = h^*_0(x^*,y^*) - (1-e^{-\alpha^* t^*})\zeta_{OK}^*(x^*,y^*).
\end{equation*}
The parameter $\alpha^*$ is related to the characteristic time $t^*_0$ under consideration. We chose
$$
  1-e^{-\alpha^* t^*_0} = \frac23 \Leftrightarrow \alpha^* = \frac{\log 3}{t^*_0}.
$$
The various parameters used in the computations are given in Table \ref{tab:TotalEnergy}.

Other time laws are possible and we refer to 
\cite{Dutykh2006} for more details. In the present numerical computations 
we chose $h^*_0(x^*,y^*) = h_0 =$ const. This choice is not only made for the sake of simplicity.
Another reason is that Okada's solution is derived within the assumption of an elastic half-space which does not
take into account the bathymetry. In order to be coherent with this solution we assume 
the bottom to be flat before deformation.

\begin{table}
     \begin{center}
       \begin{tabular}{c|l}
         \hline\hline
         \textit{Parameter} & \textit{Value} \\
         \hline
         Dip angle, $\delta$ & $13^\circ$ \\
         \hline
         Slip angle, $\theta$ & $90^\circ$ \\
         \hline
         Fault length, $L^*$ & $18$ km \\
         \hline
         Fault width, $W^*$ & $14$ km \\
         \hline
         Fault depth & $5$ km \\
         \hline
         Slip along the fault & $10$ m \\
         \hline
         Poisson ratio & $0.27$ \\
         \hline
         Young modulus & $9.5\times 10^{9}$ Pa \\
         \hline
         Acceleration due to gravity, $g$ & $9.81$ m/$s^2$ \\
         \hline
         Water depth, $h_0$ & $1$ km \\
         \hline
         Characteristic rise time, $t^*_0$ & $8$ s \\
         \hline
         $\alpha^* = \log(3)/t^*_0$ & $0.1373$ $s^{-1}$ \\
         \hline\hline
       \end{tabular}
       \caption{Values of physical parameters used for the energy density computations.}
       \label{tab:TotalEnergy}
     \end{center}
\end{table}

\begin{figure}
	\centering
		\subfigure[Free surface]%
		{\includegraphics[width=0.32\textwidth]{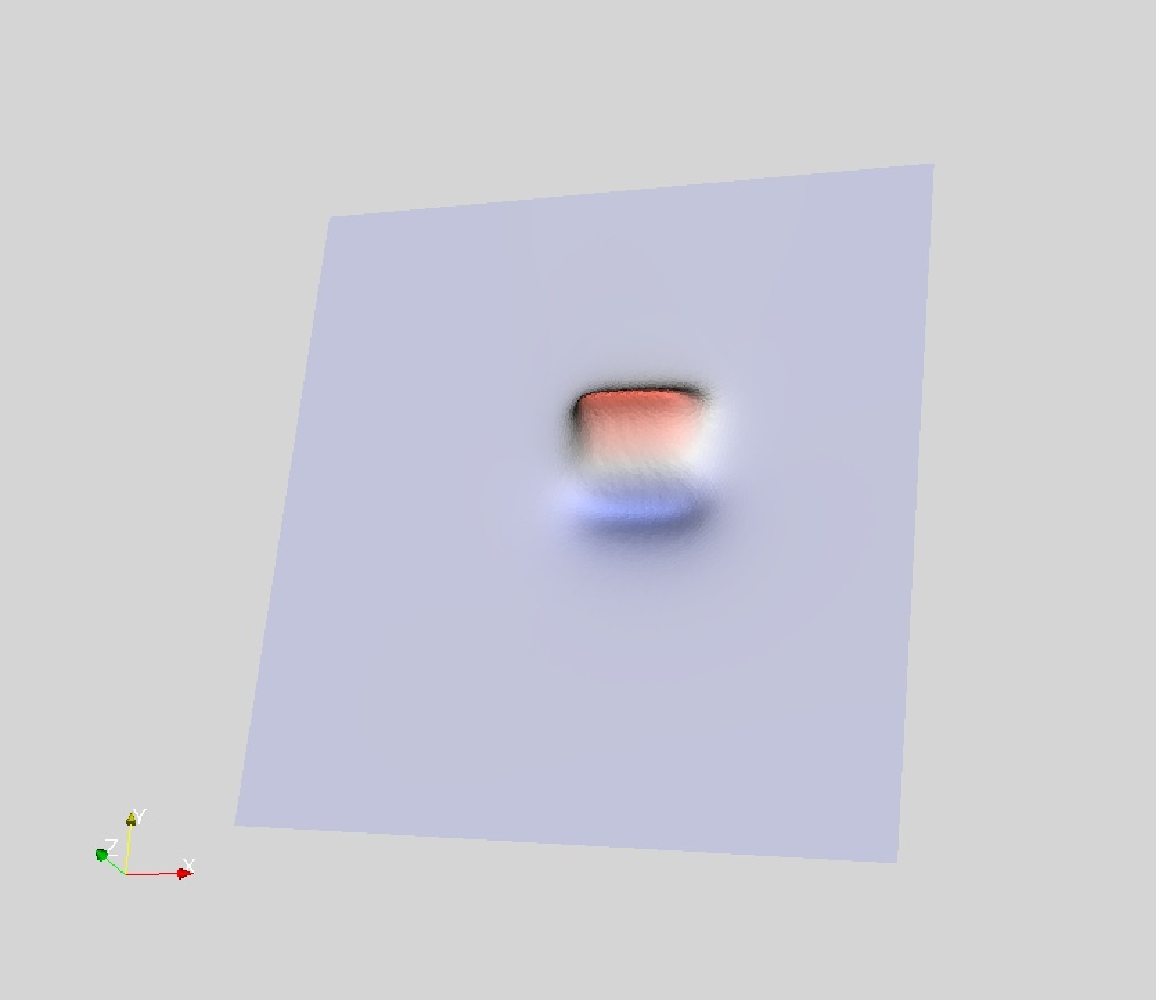}}
		\subfigure[Total energy]%
		{\includegraphics[width=0.32\textwidth]{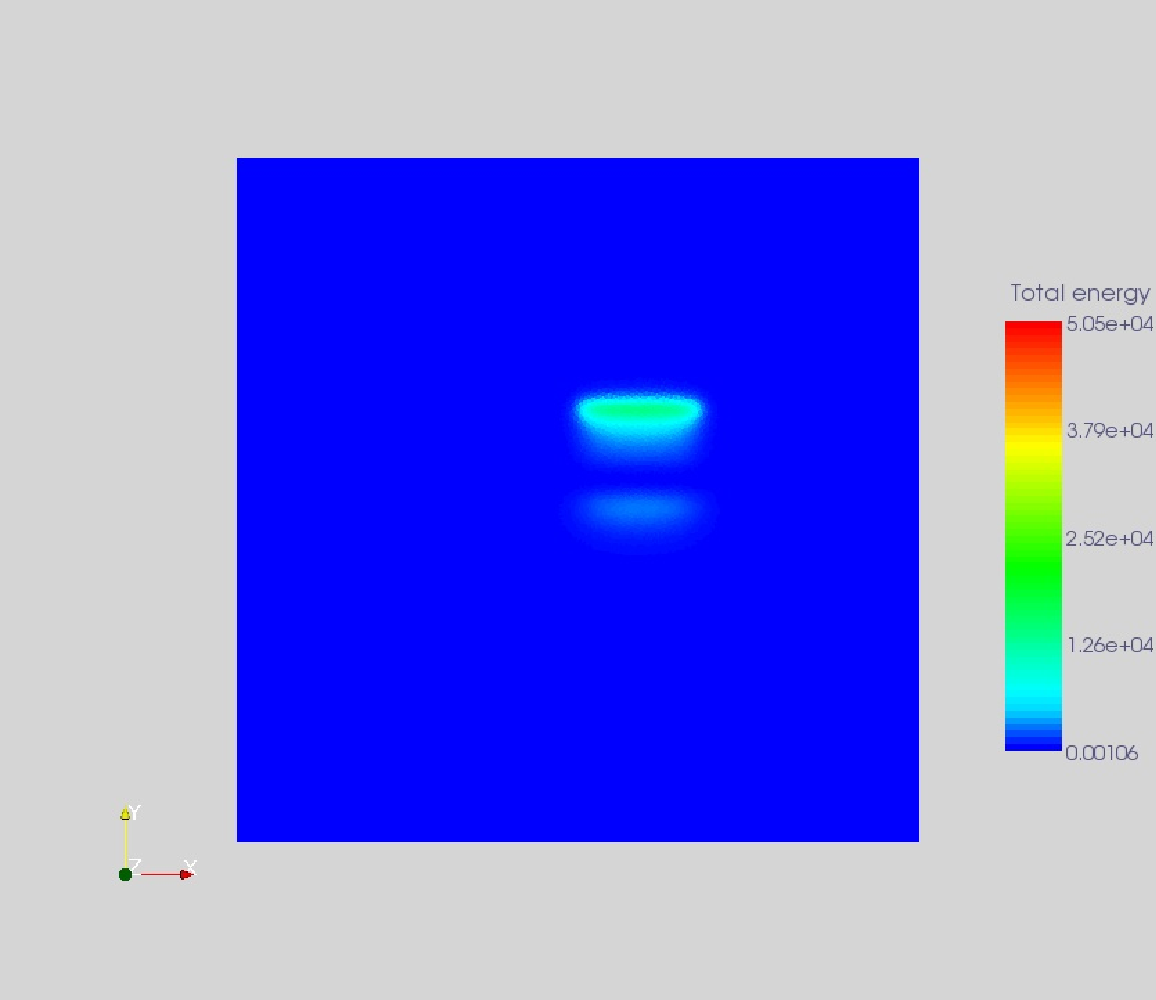}}
		\subfigure[Potential energy]%
		{\includegraphics[width=0.32\textwidth]{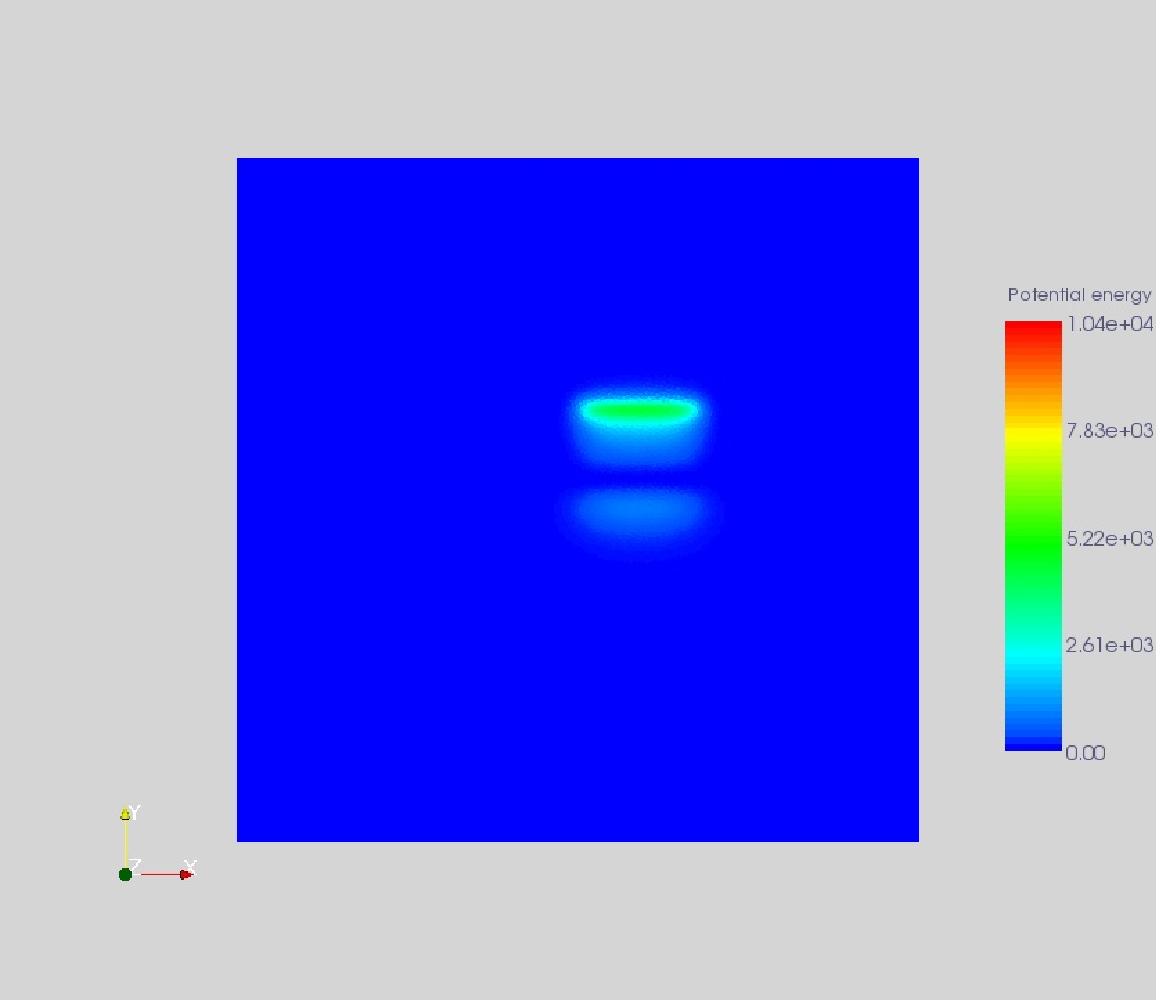}}
		\caption{Tsunami generation leading to a dipolar wave form; $t^* = 4$ s. The same scales are used in
Figures \ref{fig:freesurf3}--\ref{fig:freesurf40}: for the free surface, red denotes a rise of sea level; 
for the total energy, the scale goes from 0 (blue) to $5 \times 10^4$ (red); for the potential energy, the scale goes from 0 (blue) to $10^4$ (red).}
	\label{fig:freesurf3}
\end{figure}

\begin{figure}
	\centering
		\subfigure[Free surface]%
		{\includegraphics[width=0.32\textwidth]{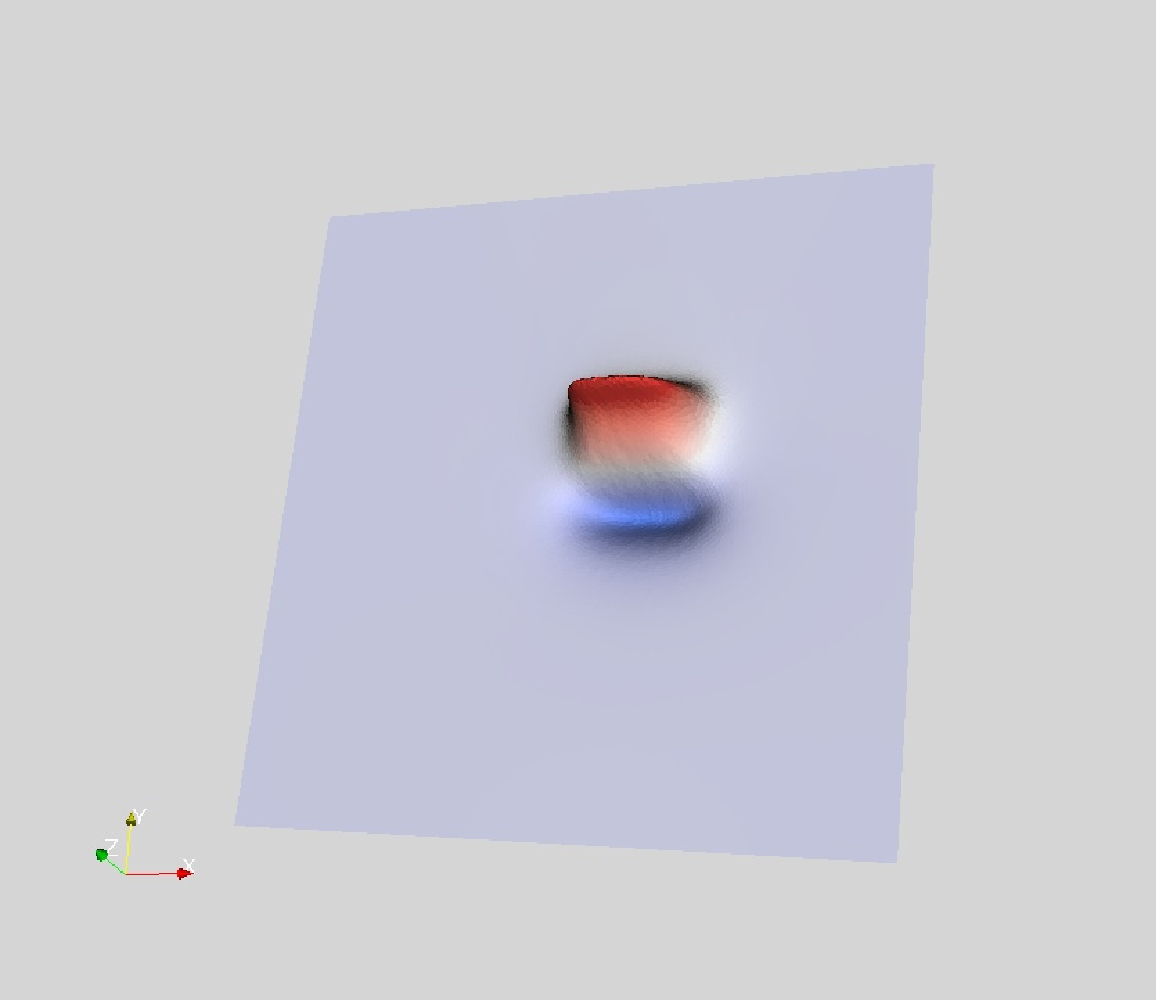}}
		\subfigure[Total energy]%
		{\includegraphics[width=0.32\textwidth]{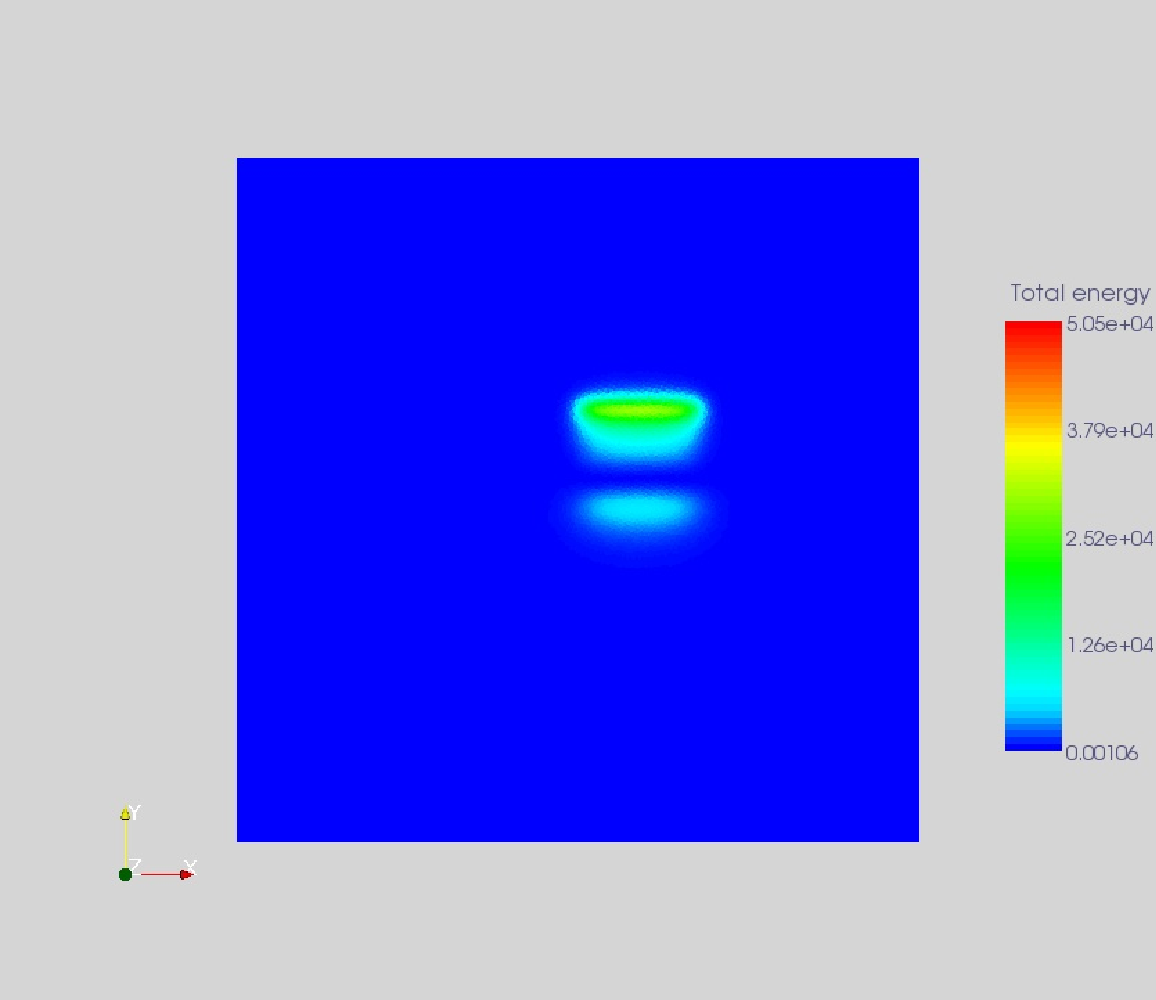}}
		\subfigure[Potential energy]%
		{\includegraphics[width=0.32\textwidth]{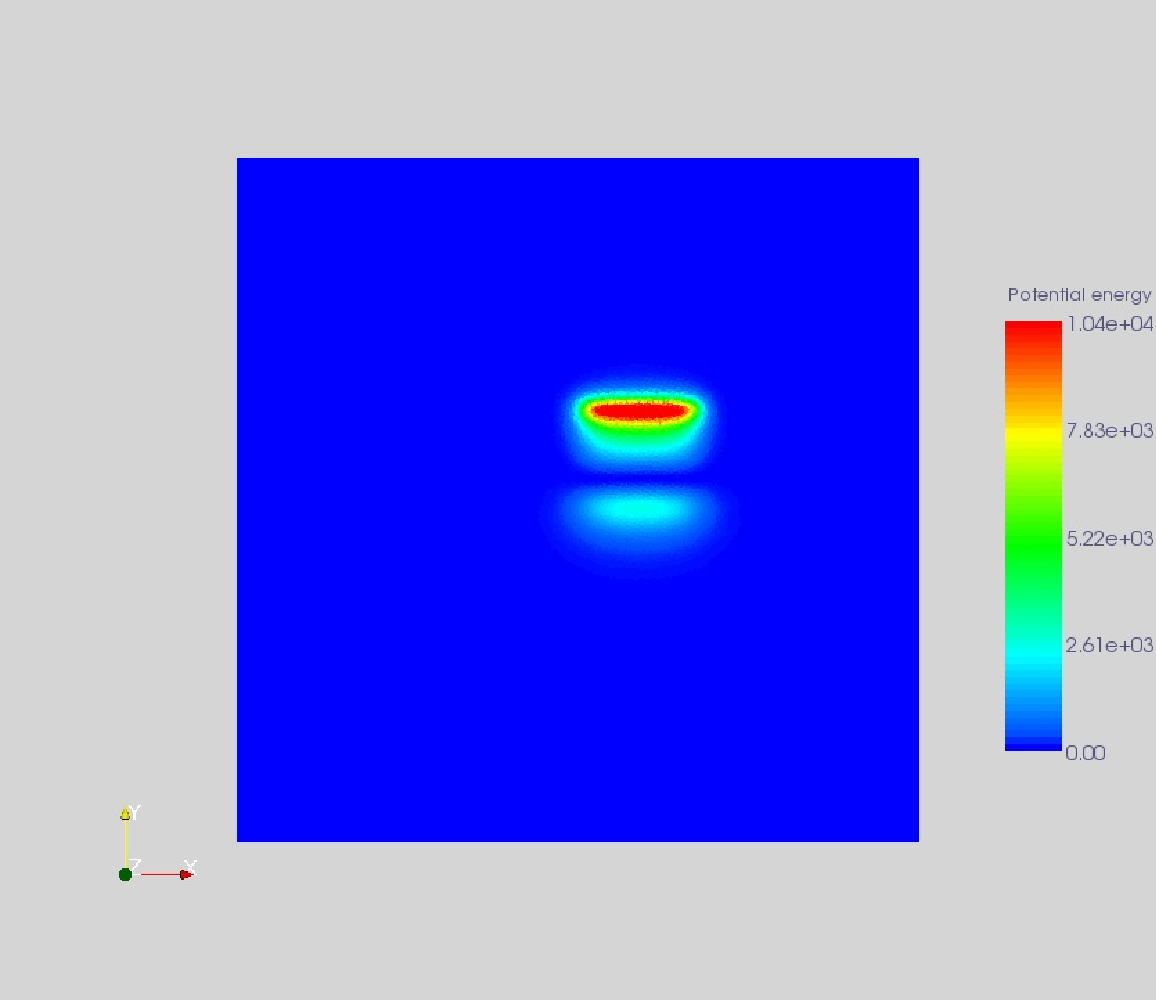}}
		\caption{$t^* = 6$ s}
	\label{fig:freesurf5}
\end{figure}

\begin{figure}
	\centering
		\subfigure[Free surface]%
		{\includegraphics[width=0.32\textwidth]{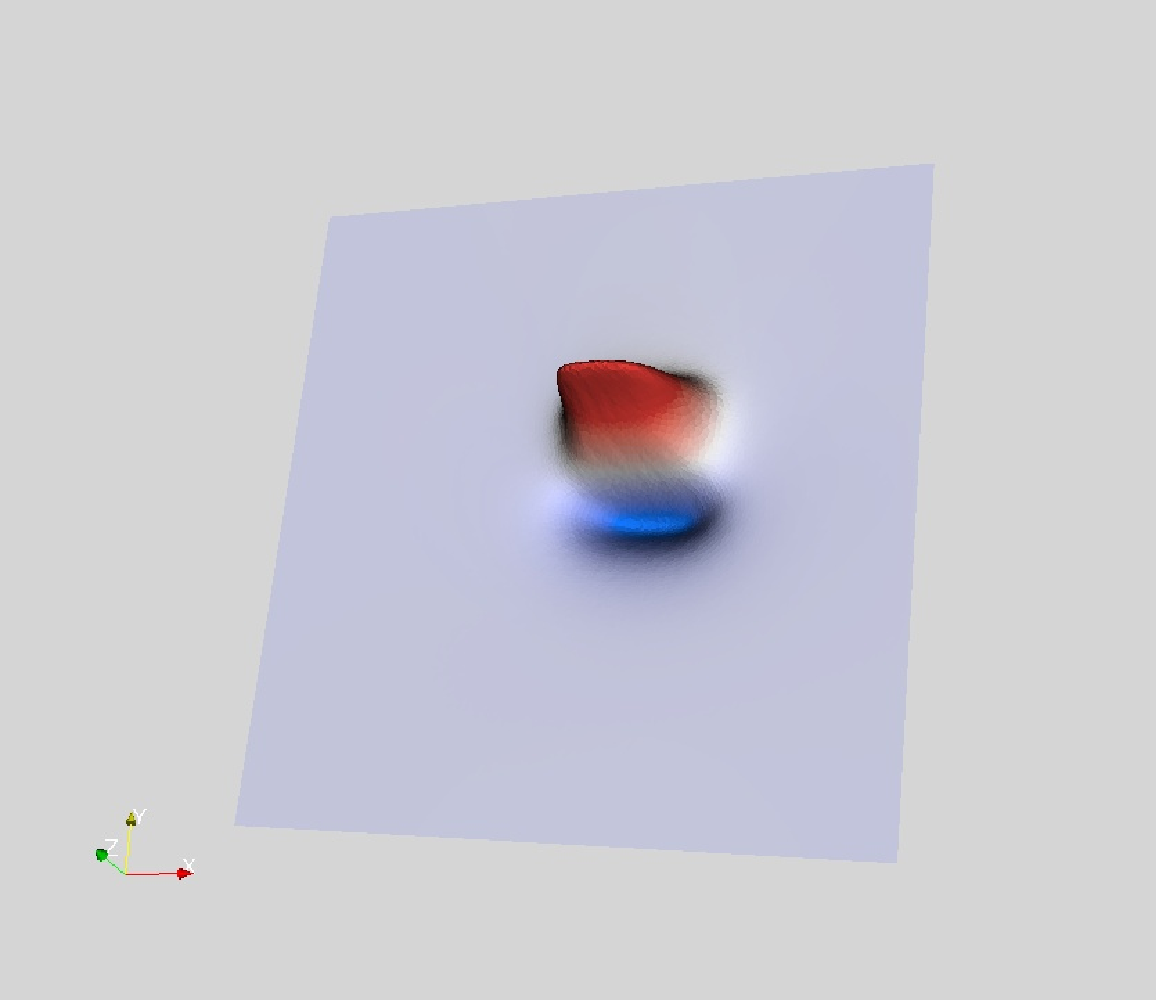}}
		\subfigure[Total energy]%
		{\includegraphics[width=0.32\textwidth]{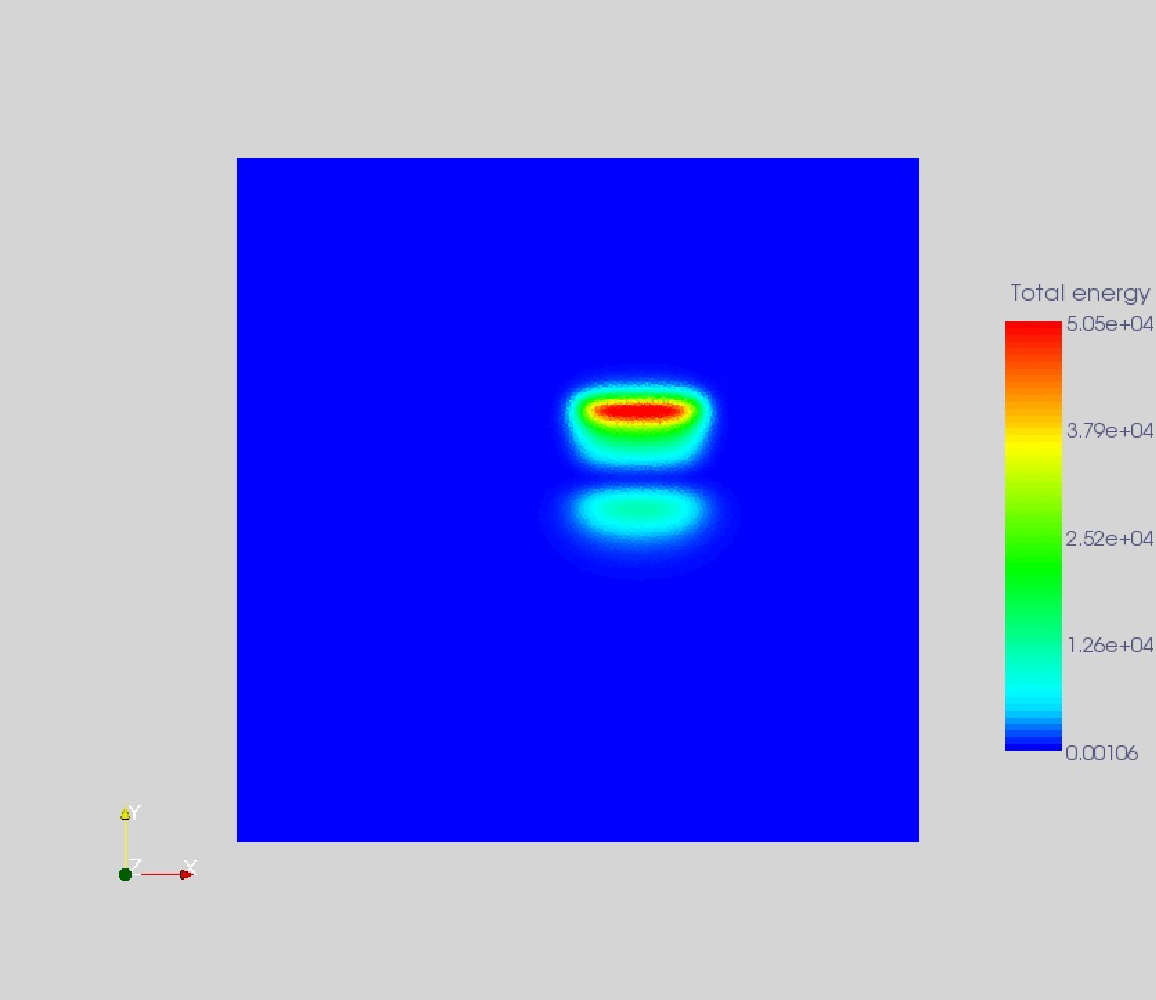}}
		\subfigure[Potential energy]%
		{\includegraphics[width=0.32\textwidth]{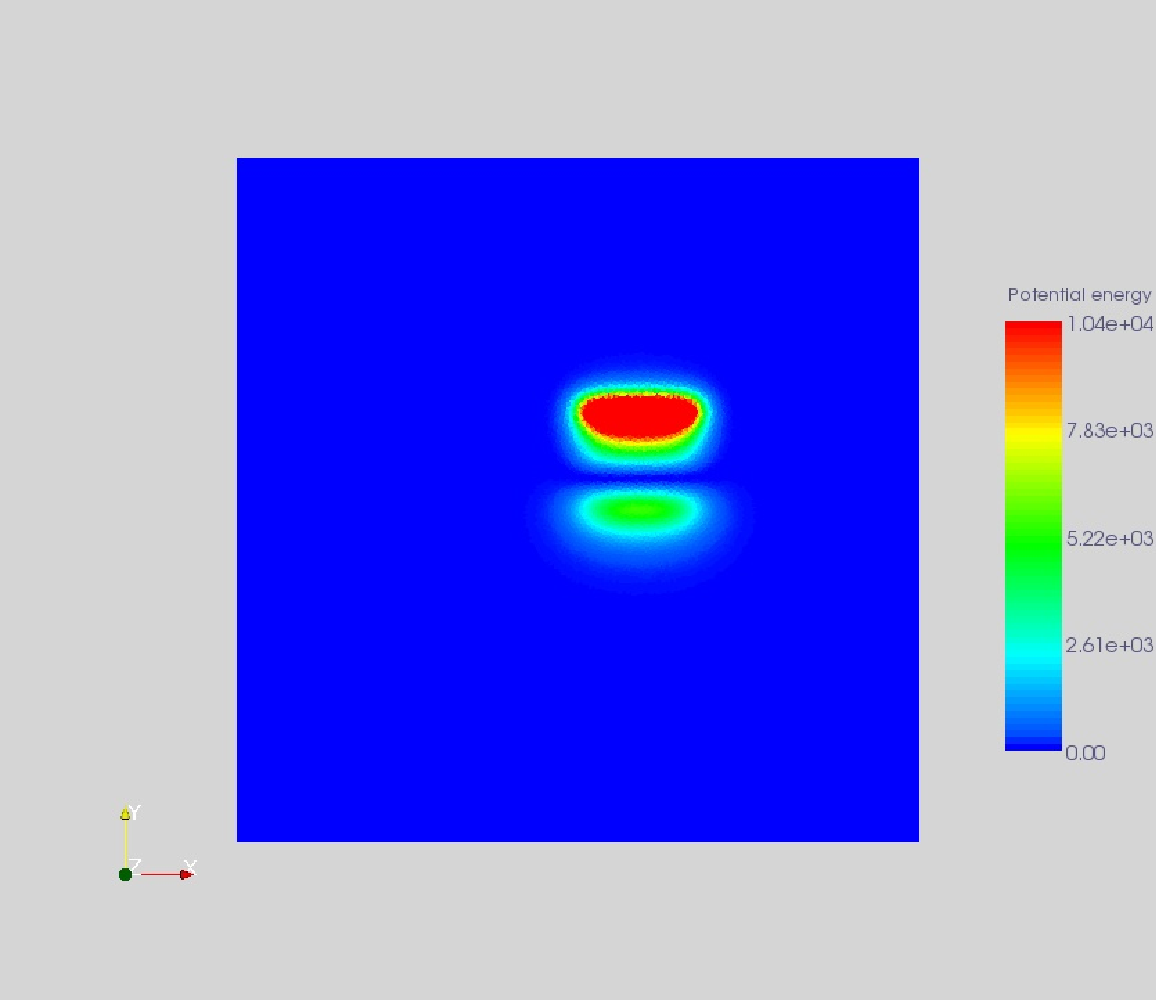}}
		\caption{$t^* = 10$ s}
	\label{fig:freesurf8}
\end{figure}

\begin{figure}
	\centering
		\subfigure[Free surface]%
		{\includegraphics[width=0.32\textwidth]{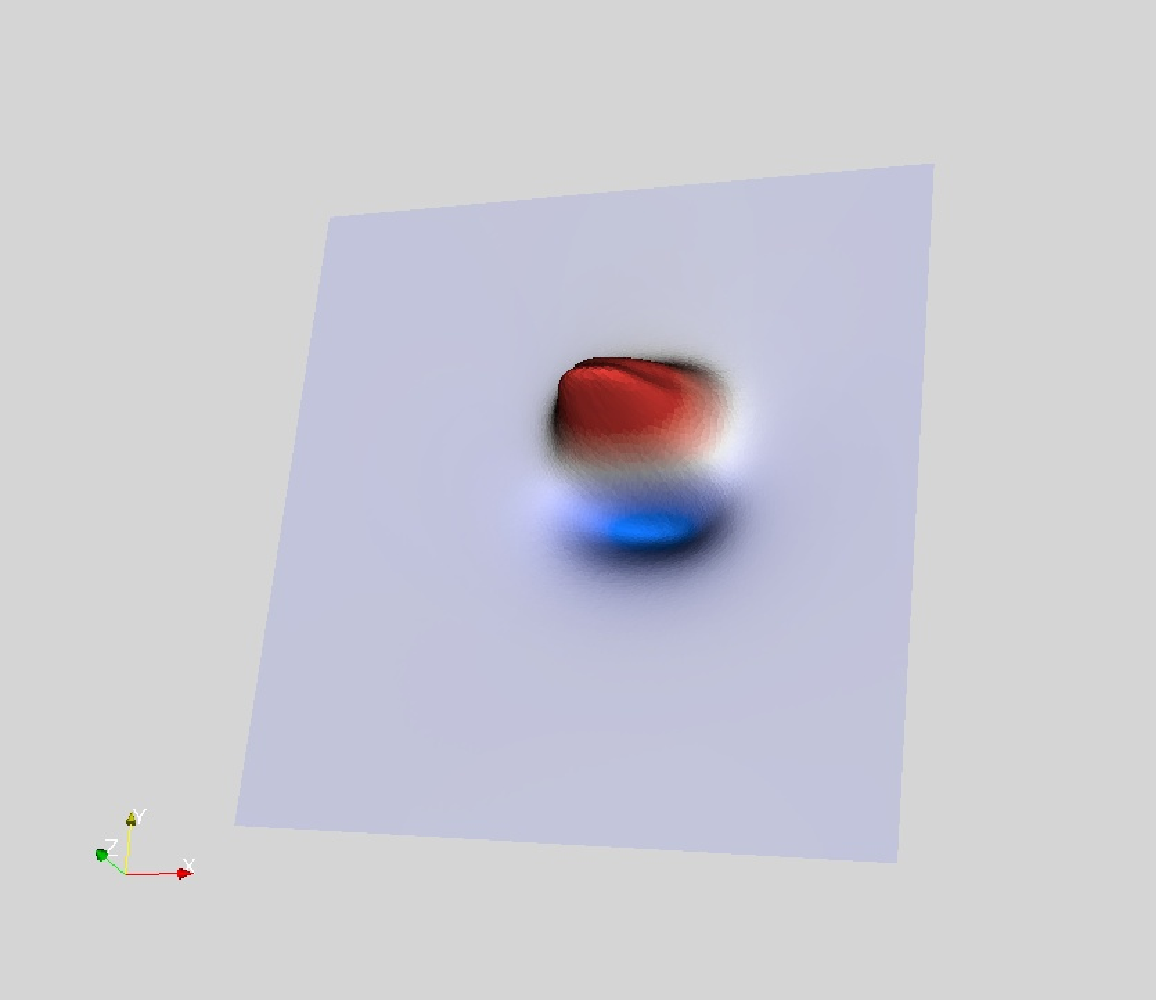}}
		\subfigure[Total energy]%
		{\includegraphics[width=0.32\textwidth]{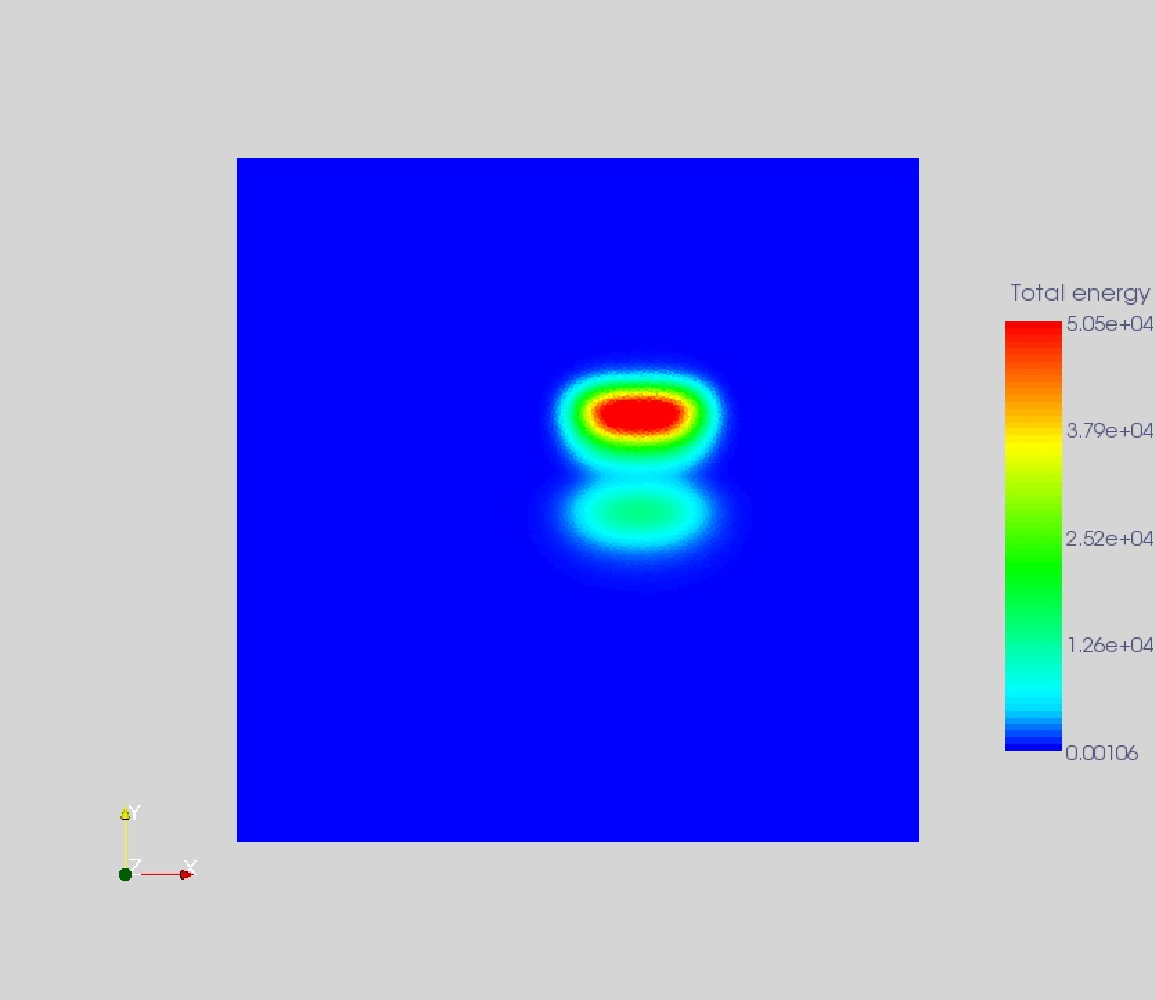}}
		\subfigure[Potential energy]%
		{\includegraphics[width=0.32\textwidth]{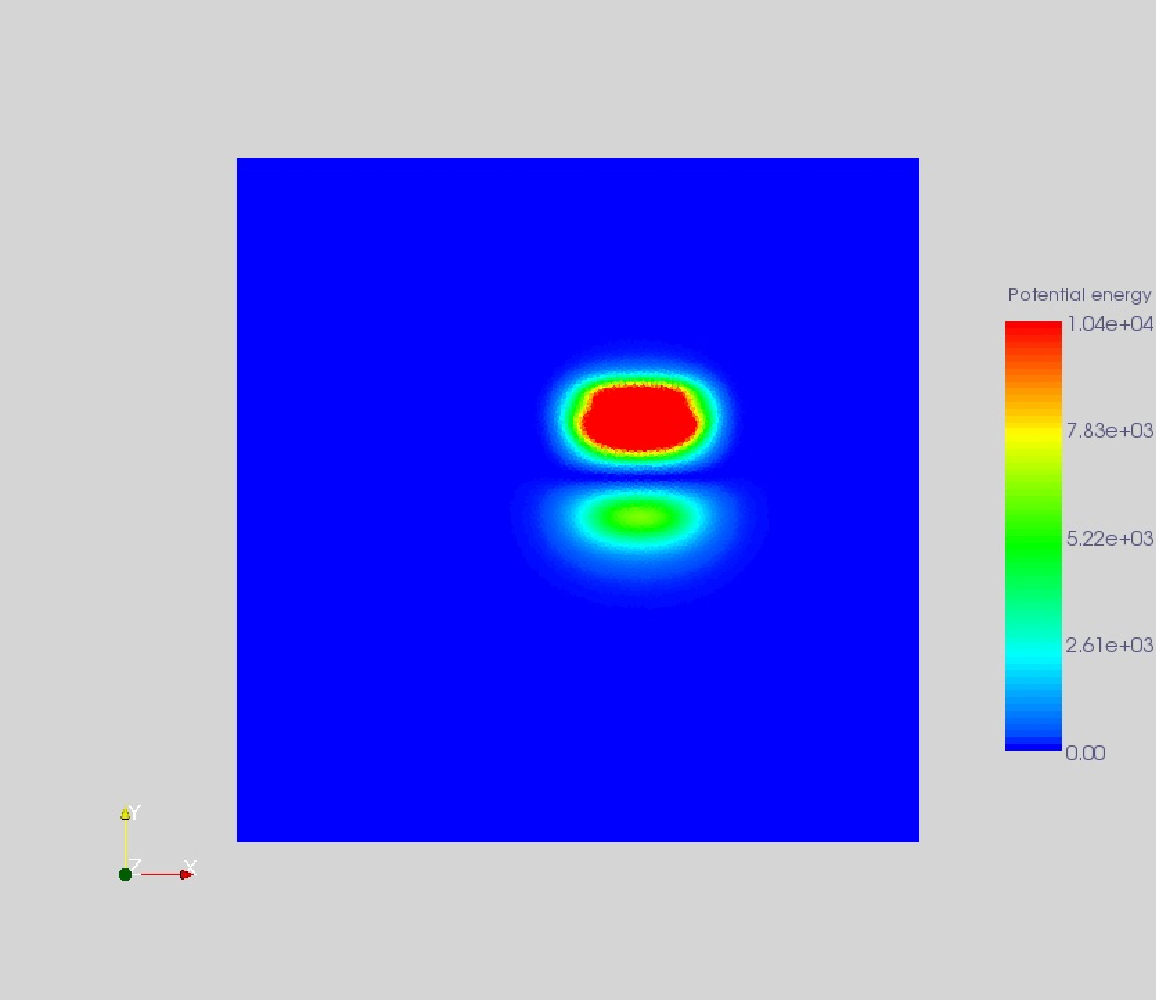}}
		\caption{$t^* = 20$ s}
	\label{fig:freesurf10}
\end{figure}

\begin{figure}
	\centering
		\subfigure[Free surface]%
		{\includegraphics[width=0.32\textwidth]{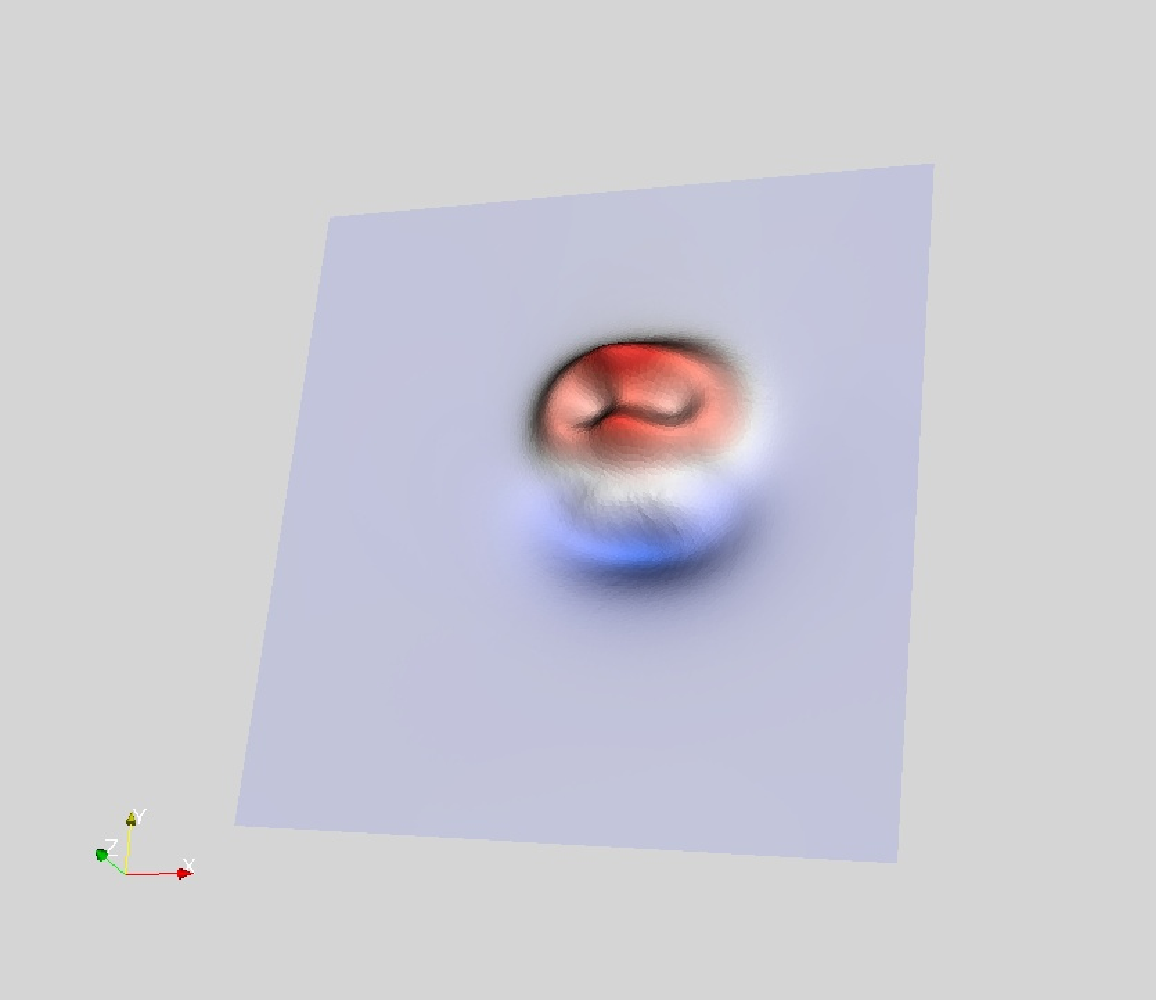}}
		\subfigure[Total energy]%
		{\includegraphics[width=0.32\textwidth]{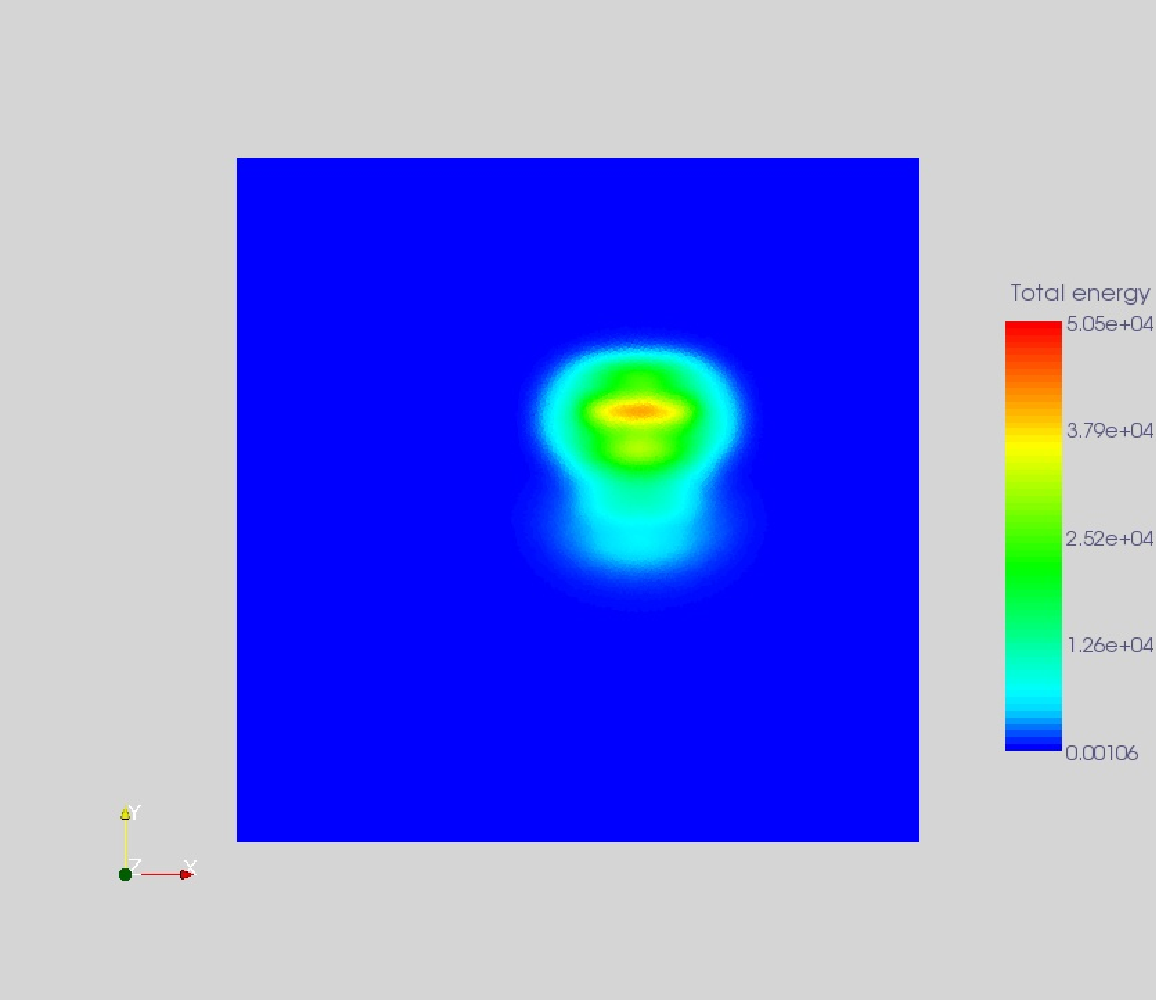}}
		\subfigure[Potential energy]%
		{\includegraphics[width=0.32\textwidth]{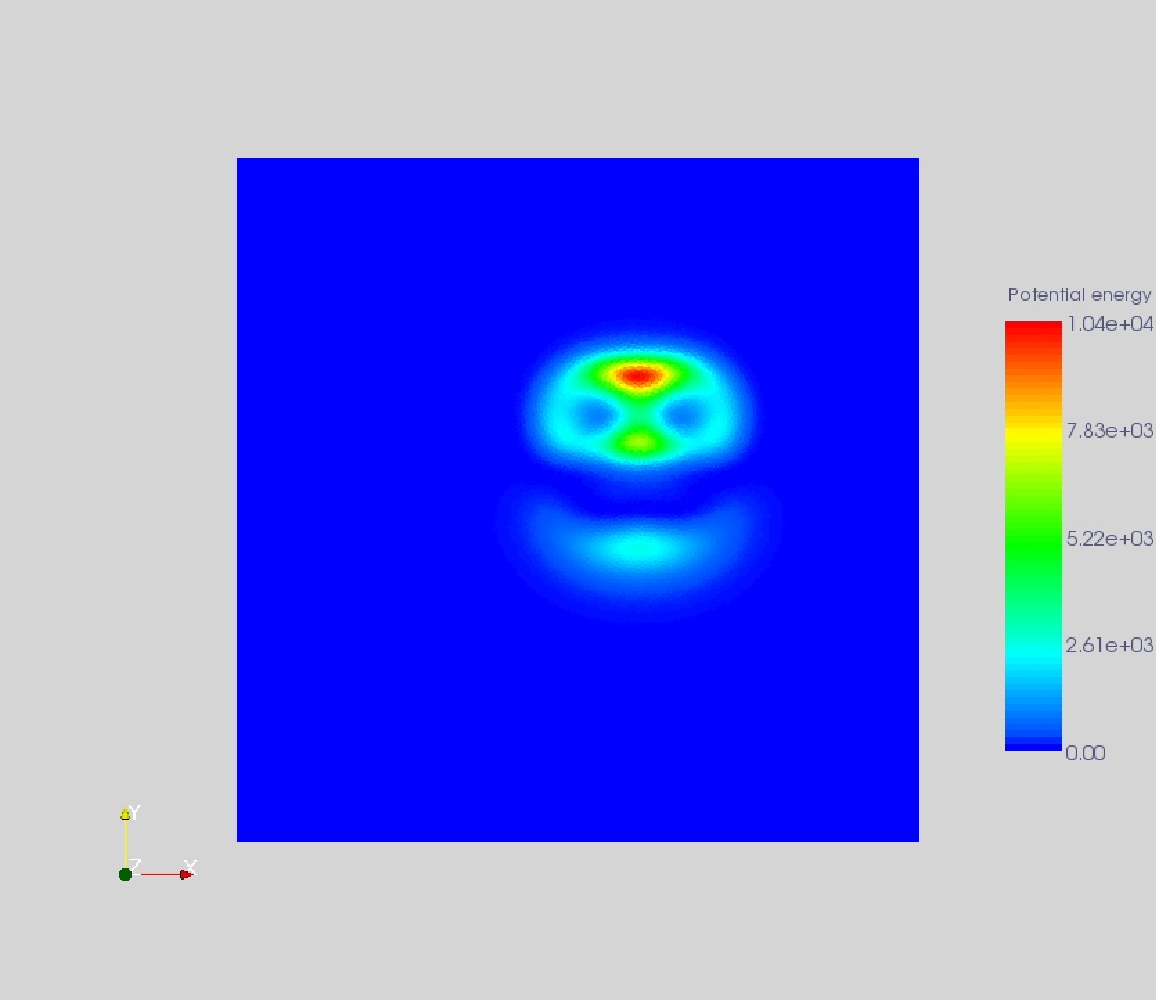}}
		\caption{$t^* = 40$ s}
	\label{fig:freesurf15}
\end{figure}

\begin{figure}
	\centering
		\subfigure[Free surface]%
		{\includegraphics[width=0.32\textwidth]{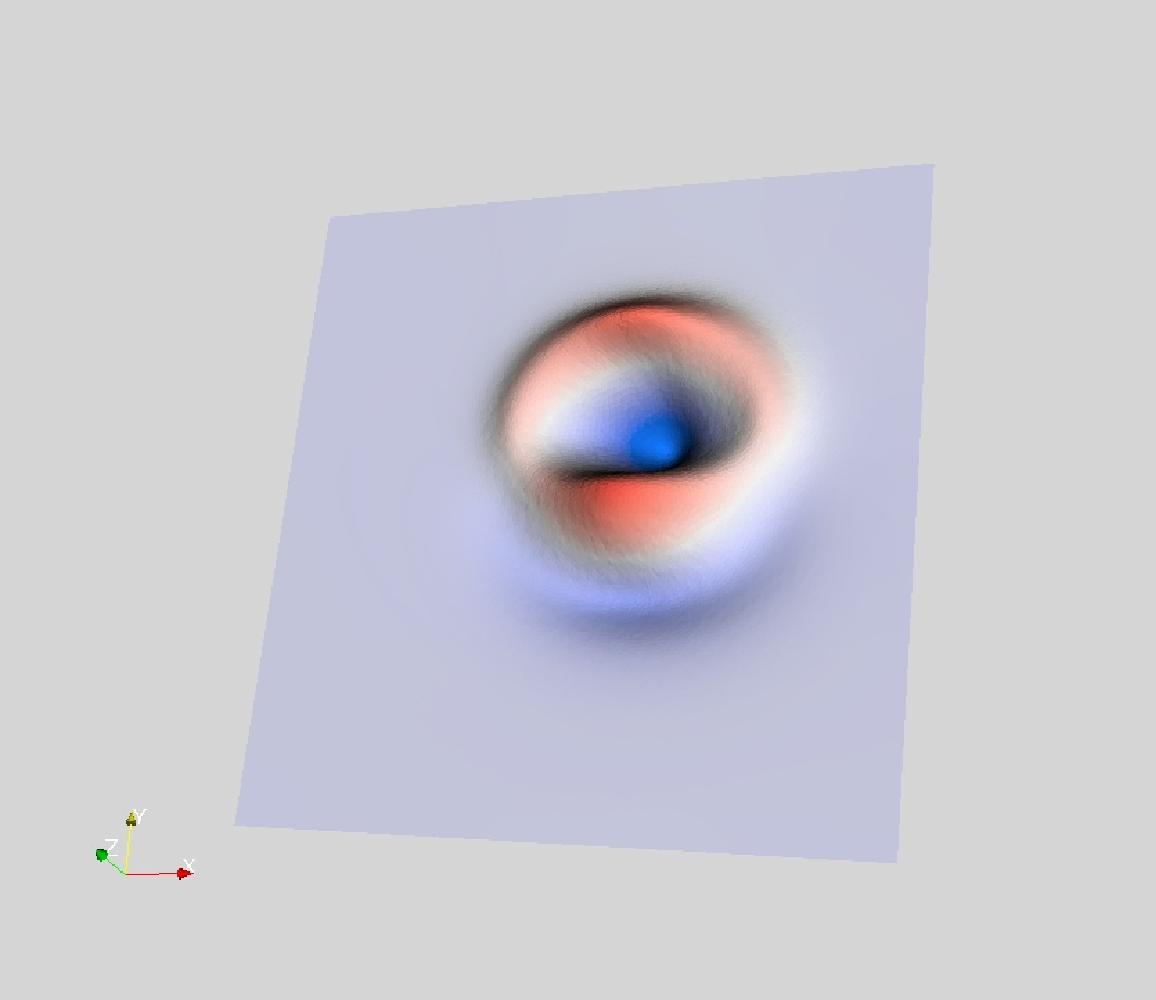}}
		\subfigure[Total energy]%
		{\includegraphics[width=0.32\textwidth]{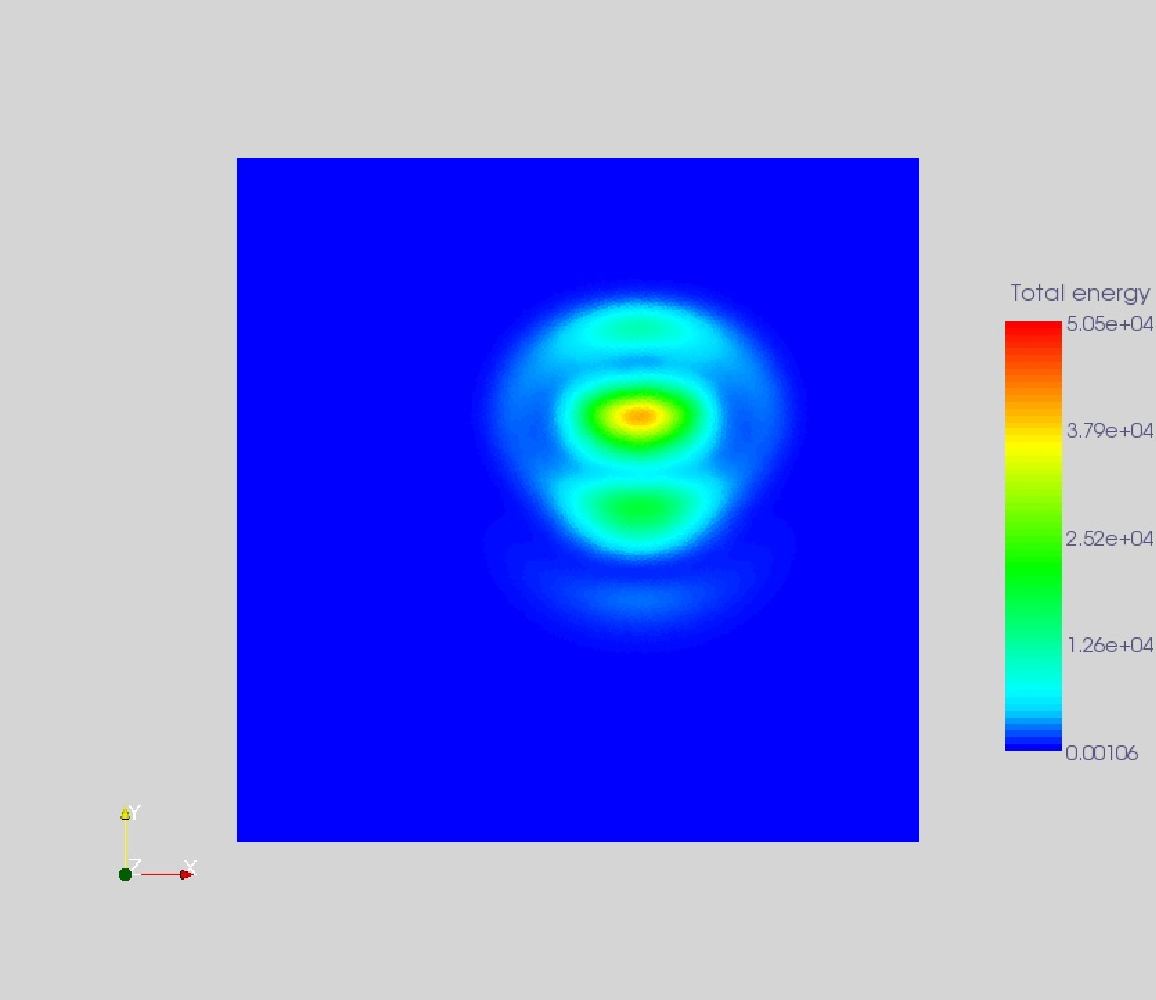}}
		\subfigure[Potential energy]%
		{\includegraphics[width=0.32\textwidth]{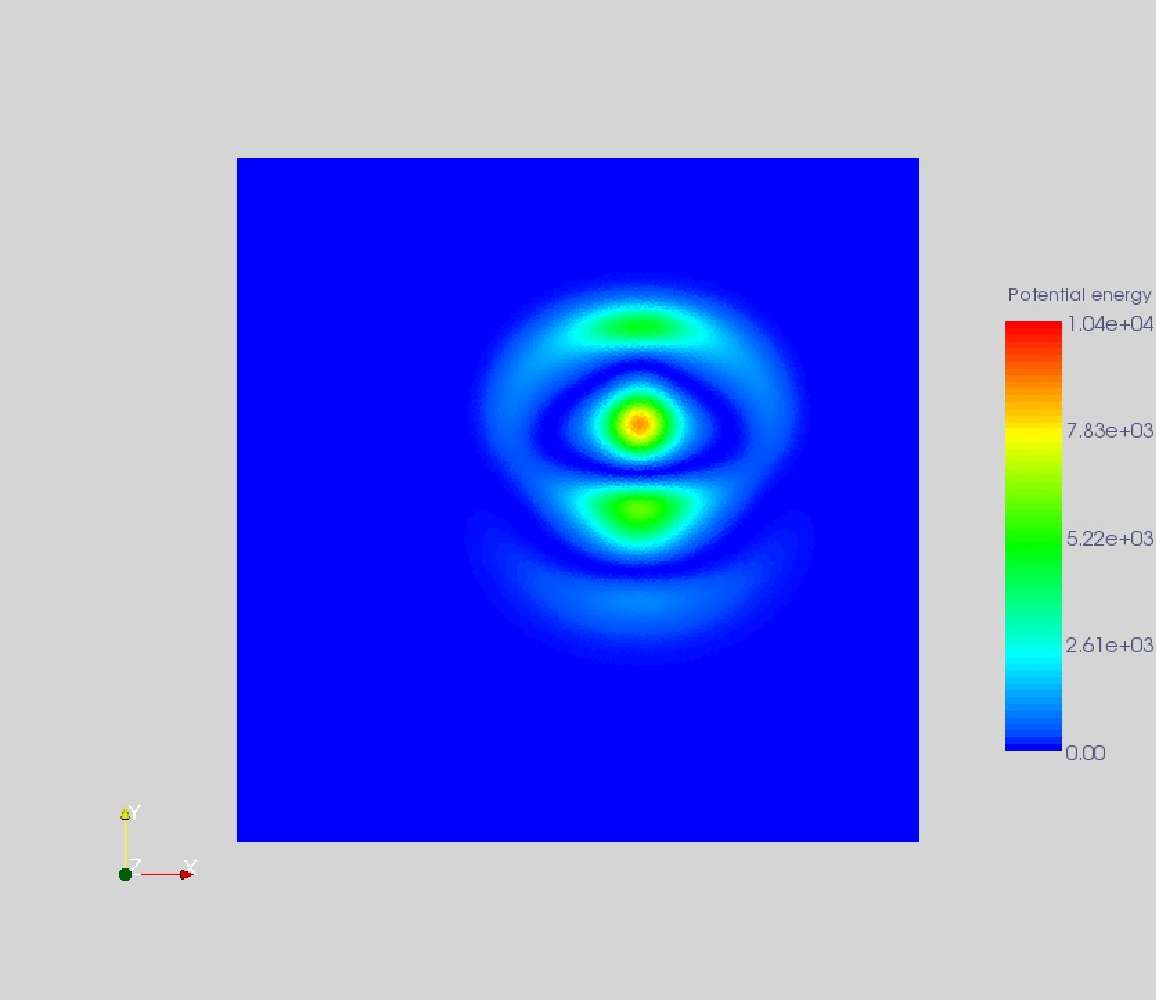}}
		\caption{$t^* = 80$ s}
	\label{fig:freesurf20}
\end{figure}

\begin{figure}
	\centering
		\subfigure[Free surface]%
		{\includegraphics[width=0.32\textwidth]{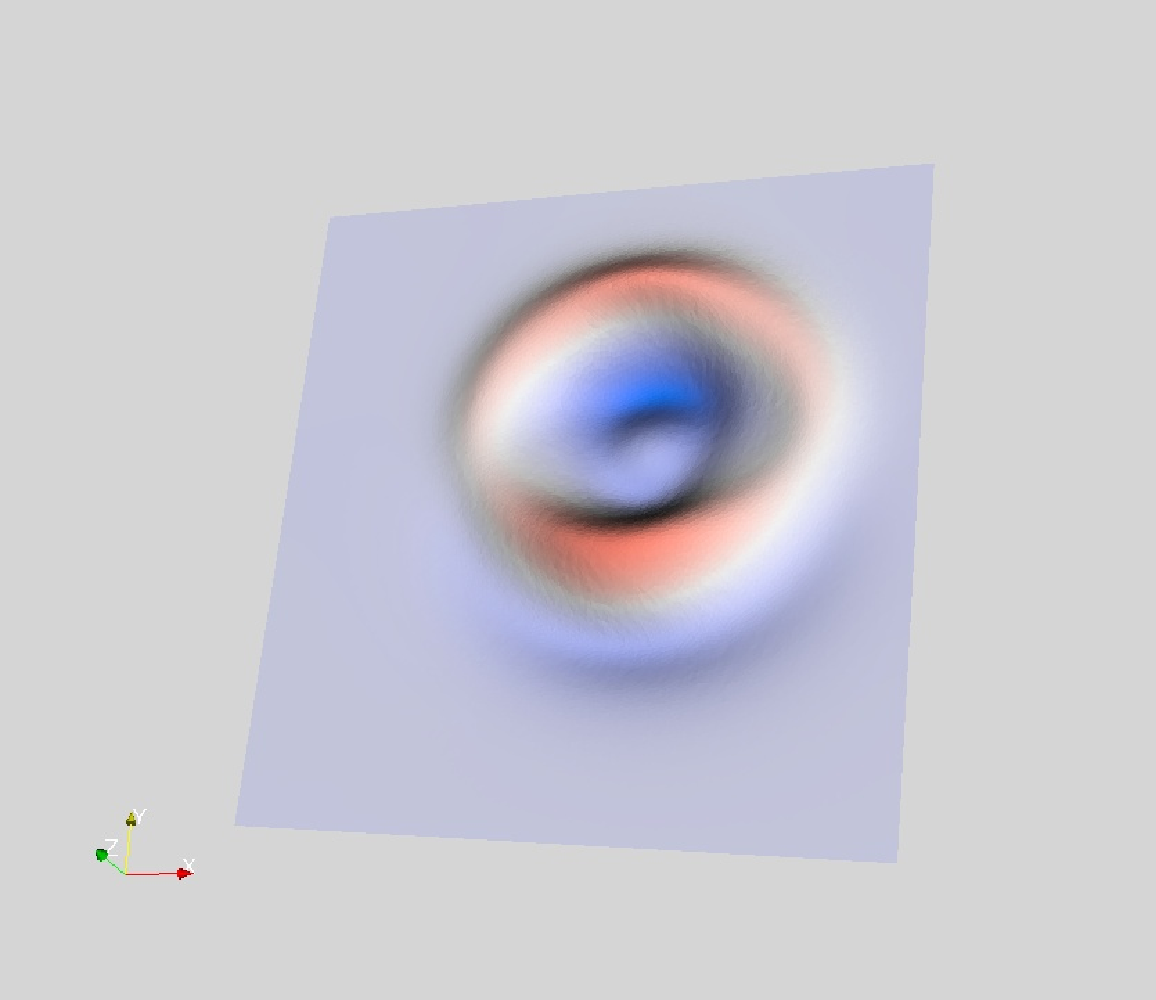}}
		\subfigure[Total energy]%
		{\includegraphics[width=0.32\textwidth]{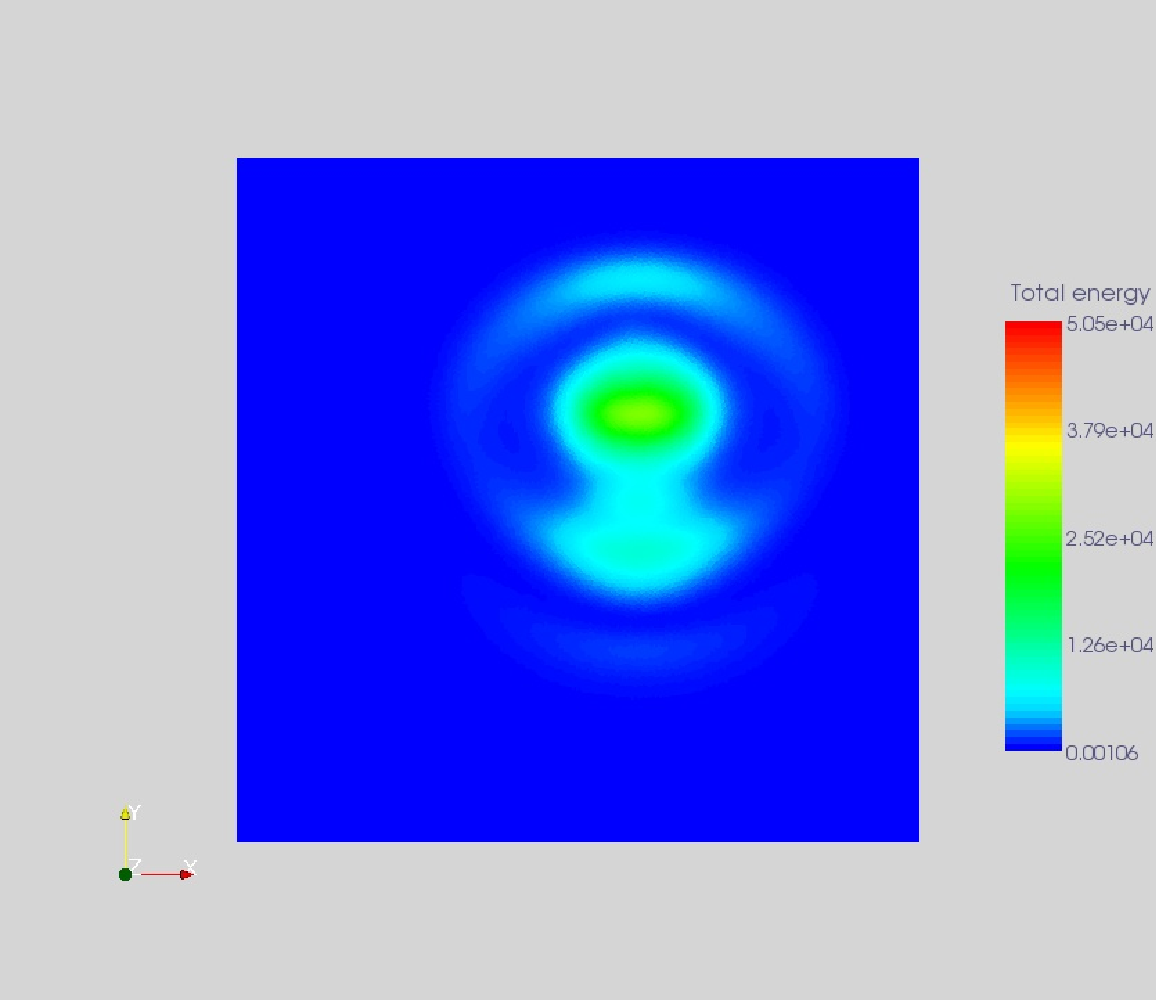}}
		\subfigure[Potential energy]%
		{\includegraphics[width=0.32\textwidth]{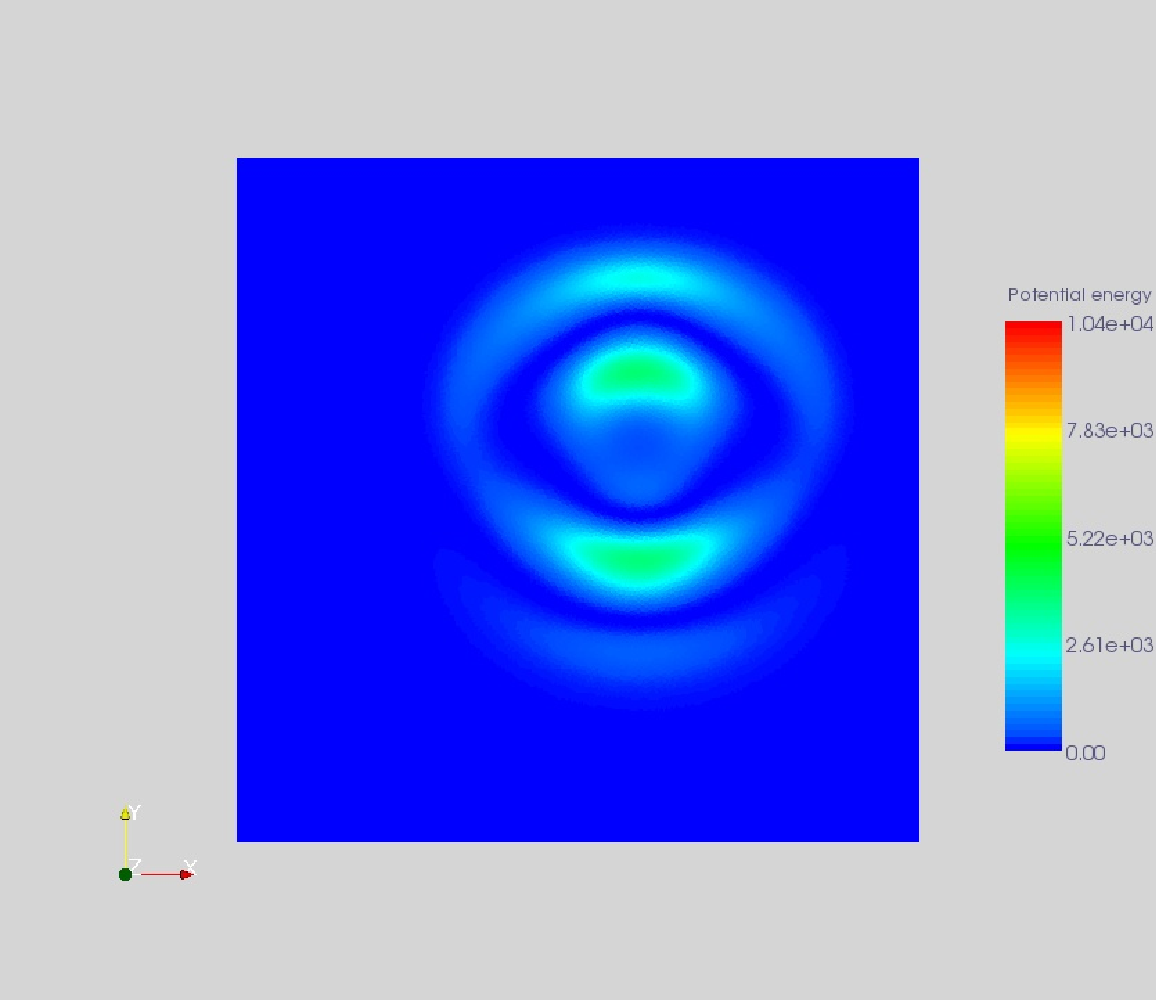}}
		\caption{$t^* = 120$ s}
	\label{fig:freesurf30}
\end{figure}

\begin{figure}
	\centering
		\subfigure[Free surface]%
		{\includegraphics[width=0.32\textwidth]{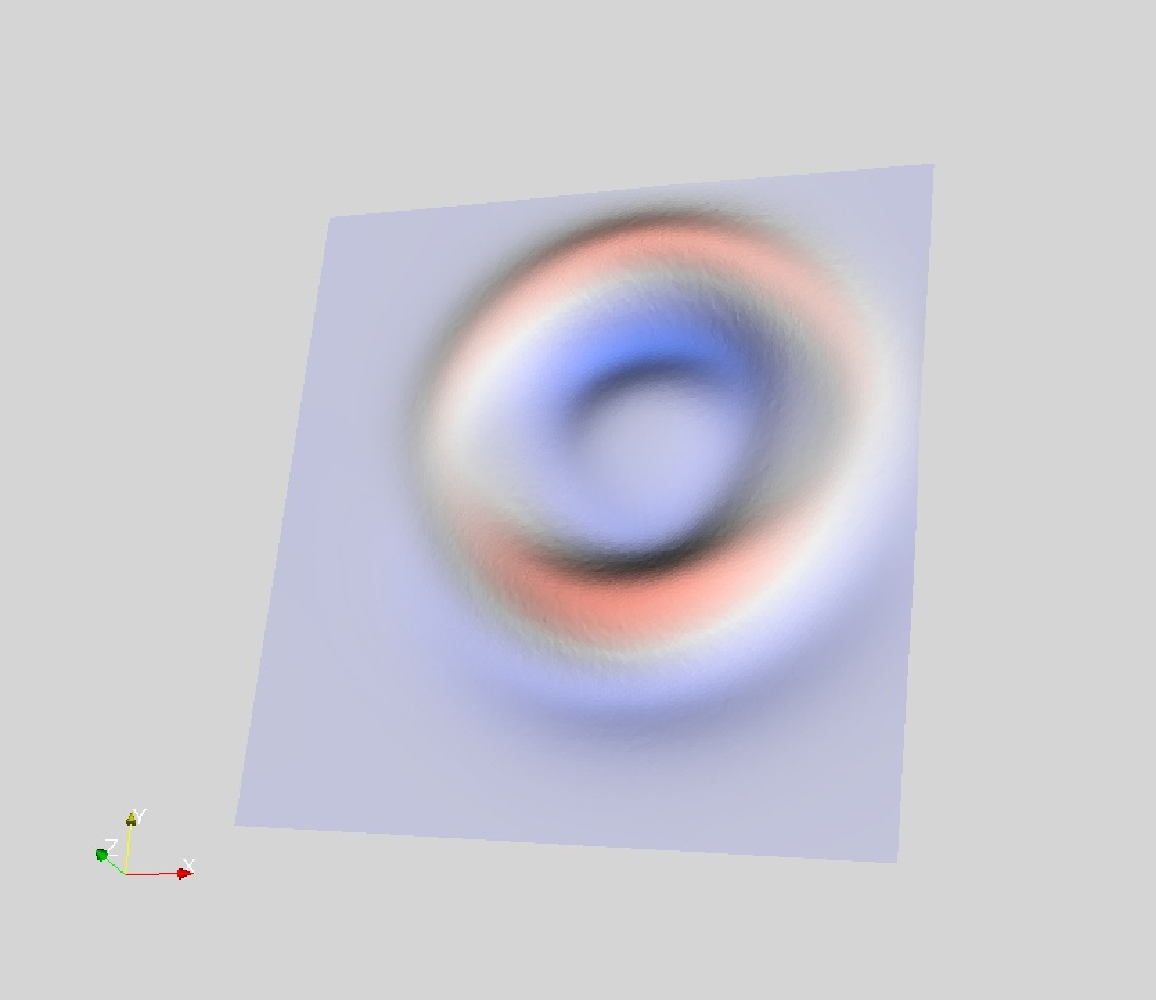}}
		\subfigure[Total energy]%
		{\includegraphics[width=0.32\textwidth]{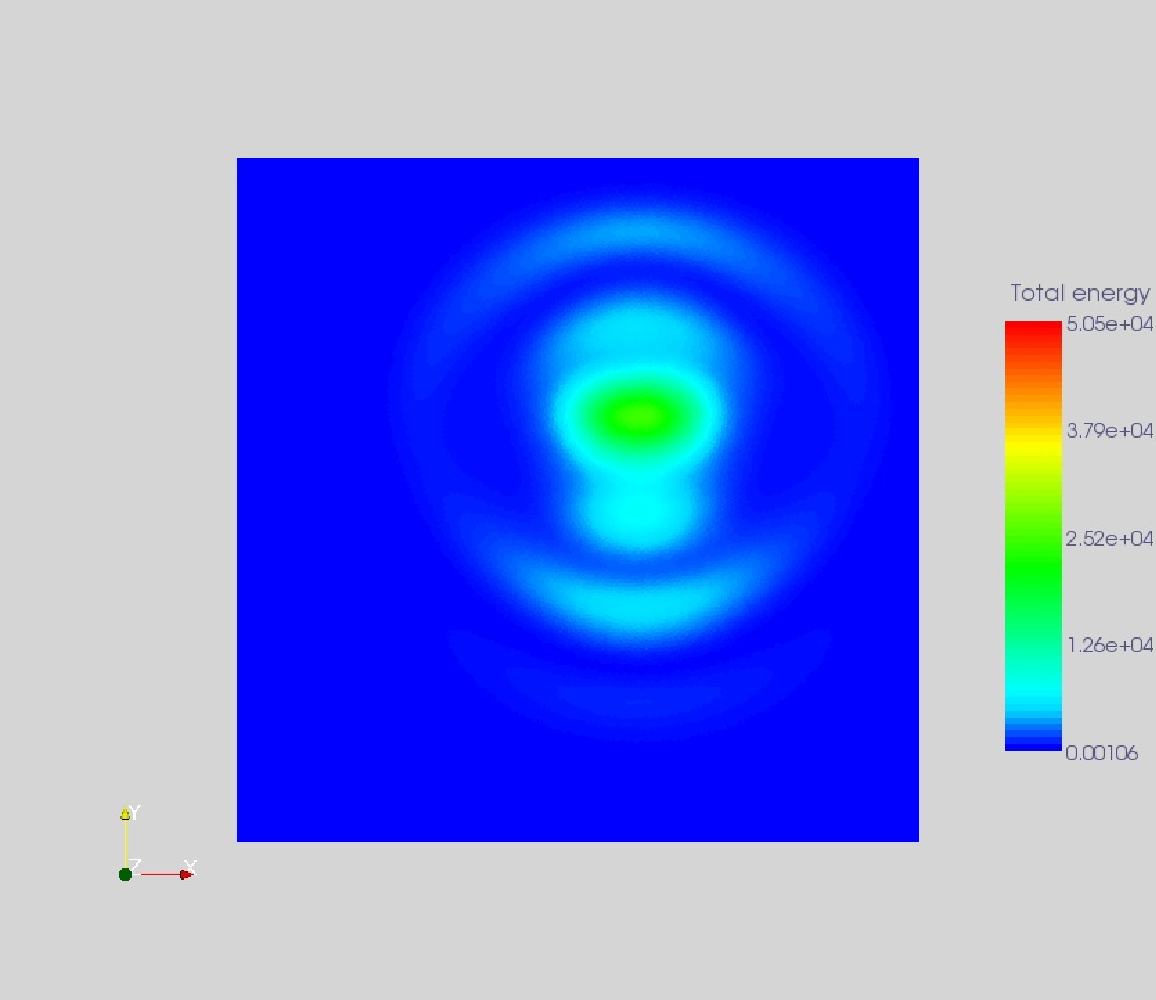}}
		\subfigure[Potential energy]%
		{\includegraphics[width=0.32\textwidth]{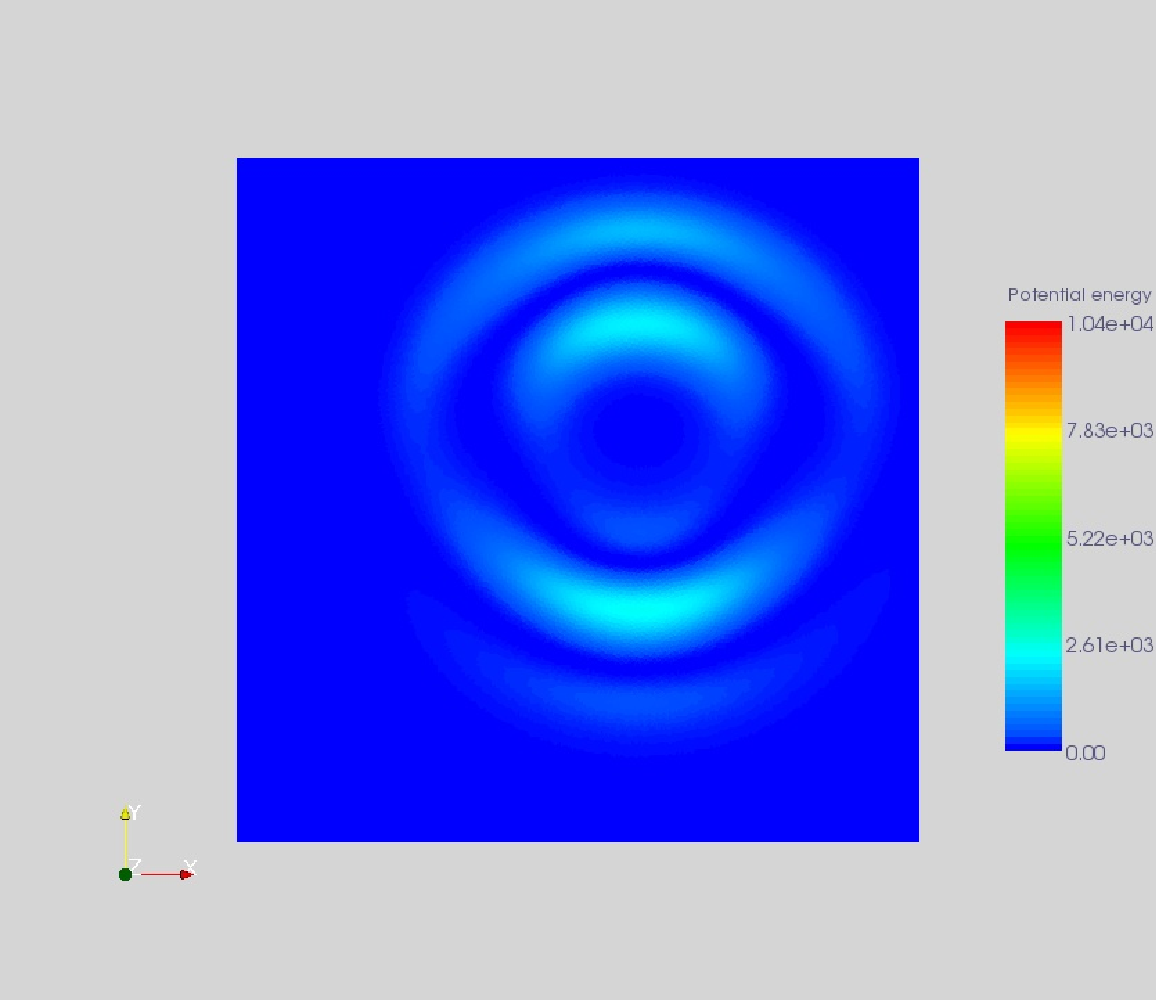}}
		\caption{$t^* = 160$ s}
	\label{fig:freesurf40}
\end{figure}

With our definition of potential energy, the total energy is nonzero both at time 
$t^*=0$ and at time $t^*\to\infty$. For practical applications it is important to
isolate the energy available to the tsunami wave. One possibility is to define the wave energy as follows:
\begin{equation}
E^*_{\rm wave} = E^* + \frac{1}{2} \rho^* g h^{*2}.
\end{equation}
Clearly $E^*_{\rm wave}=0$ both at time $t^*=0$ and at time $t^*\to\infty$ when the wave has left the generation region.
From Eq. (\ref{eq:nondispEskilsson3}), one finds that the energy equation satisfied by $E^*_{\rm wave}$ is
\begin{equation} \label{tsuenergy}
\pd{E^*_{\rm wave}}{t^*} + \nabla \cdot \left( \bar\u^*\left(E^*_{\rm wave} + \frac{1}{2}\rho^* gH^{*2} 
- \frac{1}{2}\rho^* gh^{*2} \right)\right) = \rho^* g (H^*-h^*) \pd{\zeta^*}{t^*}. 
\end{equation}
Figures \ref{fig:freesurf3}--\ref{fig:freesurf40} show the distributions of free-surface elevation $\eta^*(x^*,y^*)$, total wave energy $E_{\rm wave}^*(x^*,y^*)$ and potential wave energy $\frac{1}{2} \rho^* g \eta^{*2}$ at various times. 
One clearly sees the generation process. The formation of the leading elevation and depression waves takes a few seconds. Then
the propagation begins. As shown by \cite{Ben-M}, tsunami energy radiates primarily at right angles to a rupturing fault
(see also \cite{Kajiura1970}). The distribution of potential energy makes sense when one compares the potential energy
plots with the free-surface plots. The total energy spreads in a more uniform manner across the area affected by the waves. 

\begin{figure}
	\centering
		\subfigure[Total energy]%
		{\includegraphics[width=0.49\textwidth]{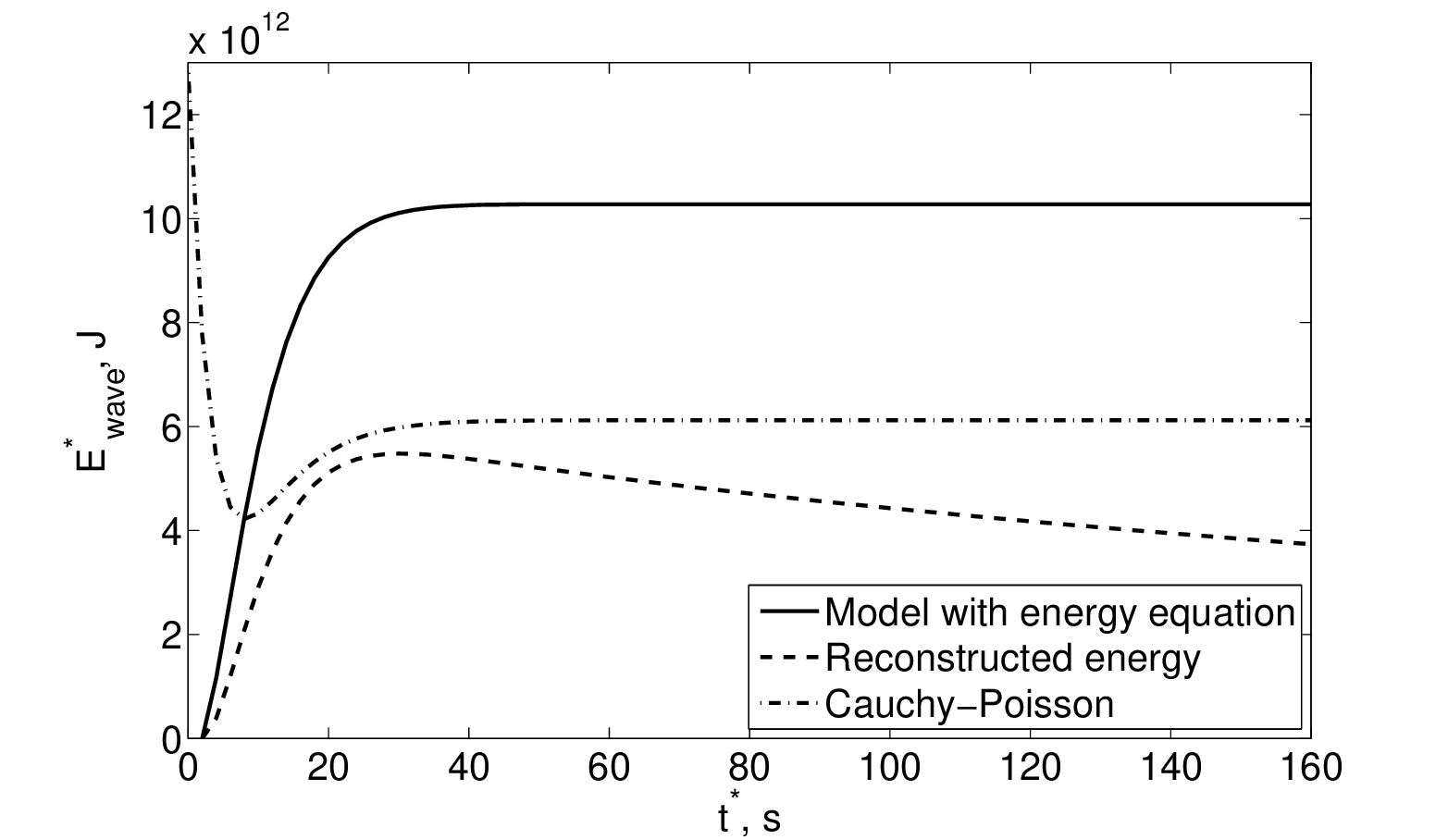}}
		\subfigure[Total, potential and kinetic energies]%
		{\includegraphics[width=0.49\textwidth]{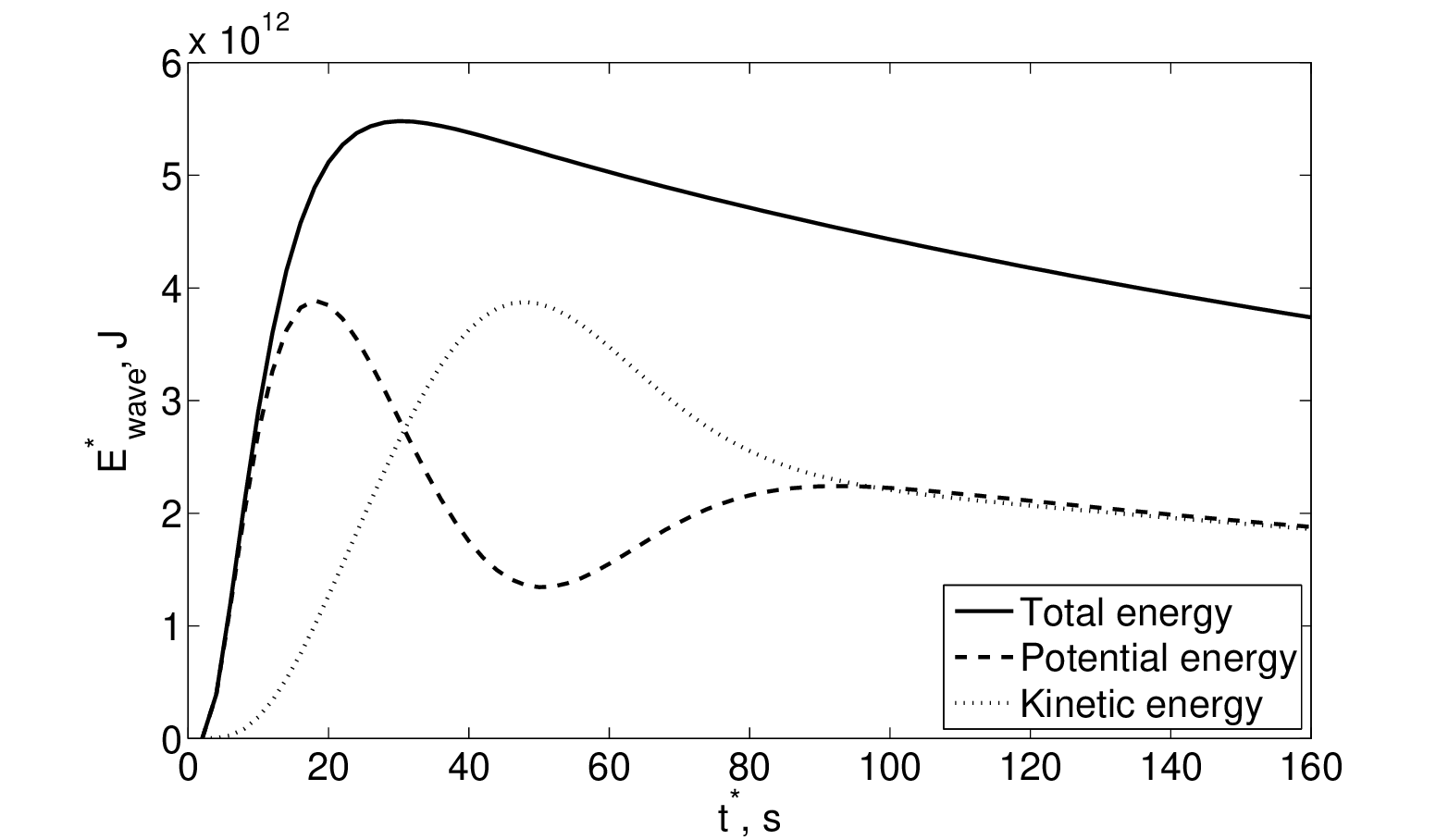}}
	\caption{(a) Total energy $\int\!\!\!\int E_{\rm wave}^* d\x^*$ as a function of time computed with the SWE with energy
(\ref{eq:nondispEskilsson1})--(\ref{eq:nondispEskilsson3}) (top curve),
the linearized water wave equations (middle curve) and
	reconstructed from the flow variables $\bar \u^*$ and $\eta^*$ when no energy equation is used (bottom curve). (b)
Partition between potential and kinetic energies as a function of time for the reconstructed energy. The top curve of (b)
is the same as the bottom curve of (a).}
	\label{fig:Energy}
\end{figure}

It is also of interest to see how the wave energy $E_{\rm wave}^*$, integrated over the whole fluid domain, varies during the
generation process. This evolution is shown in figure \ref{fig:Energy}. 
Three curves are plotted in figure \ref{fig:Energy}(a). The solid curve simply is the energy (\ref{realenergy}) integrated over the water surface. The middle curve will be explained in Section 4 (linearized theory). The bottom curve is obtained as follows. Imagine that one is solving the SWE without the energy equation. In order to look at the energy, a natural way to do it is to approximate the wave energy $E^*_{\rm wave}$ as 
\begin{equation*}
E^*_{\rm wave} \approx \frac{1}{2} \rho^* H^* |\bar \u^*|^2 + \frac{1}{2} \rho^* g \eta^{*2},
\end{equation*}
and then to integrate over the whole fluid domain.
The second main result of the present paper is that the difference between the proper way to compute the energy and 
the approximate way can be large. It is
probably due to the vertical velocity, which is completely neglected in the second approach. Once the
motion of the sea bottom has stopped, the total energy remains constant. However the reconstructed energy decreases with time.
This indicates that computing the wave energy directly from the conservative energy equation (\ref{tsuenergy}) is much better,
especially with the type of numerical method we use. In figure \ref{fig:Energy}(b), one clearly sees the exchange between
potential and kinetic energies until equipartition is reached. 

\section{Energy in the framework of the dispersive linearized equations}

In the case of tsunamis generated by earthquakes, nonlinear effects are not important during the process
of generation and propagation. This is why it is valid to use the linearized water-wave equations.
\cite{Dutykh2006a} and others showed that taking an instantaneous seabed deformation is not equivalent to instantaneously 
transferring the seabed deformation to the ocean surface, except in the framework of the linearized shallow water
equations (very long waves). The difference comes from the vertical velocities and dispersion. In this
case we must go back to the initial set of equations
(\ref{eq:EulerIncompr1})--(\ref{eq:EulerIncompr3}). Since the motion starts from the state
of rest, it can be considered as irrotational (potential flow) and one can introduce the velocity potential 
$\u^*=\nabla\phi^*$. 

We perform the linearization of the equations (\ref{eq:EulerIncompr1}) and (\ref{eq:EulerIncompr2}), and of the boundary
conditions (\ref{eq:dynbound})--(\ref{eq:kinbottom}). It is equivalent to taking the limit of the equations as $\eps \to 0$. 
For the sake of convenience, we switch back to the physical variables. The linearized
problem in dimensional variables reads \citep{Dutykh2006}
\begin{equation}\label{lapl}
  \Delta_{\perp} \phi^* + \partial^2 \phi^* \slash \partial z^{*2} = 0, \qquad (x^*,y^*,z^*) \in \mathbb{R}^2\times[-h^*, 0],
\end{equation}
\begin{eqnarray}\label{kinfreesurf}
  \pd{\phi^*}{z^*} & = & \pd{\eta^*}{t^*}, \quad z^* = 0, \quad \mbox{kinematic condition}, \\
\label{dynfreesurf}
  \pd{\phi^*}{t^*} + g\eta^* & = & 0, \quad z^* = 0, \quad \mbox{dynamic condition}.
\end{eqnarray}

Within linear theory the forces that
cause perturbations are so weak that the boundary condition
at the bottom (\ref{eq:kinbottom}) is also simplified:
\begin{equation}\label{kinsolb}
  \pd{\phi^*}{z^*} = \pd{\zeta^*}{t^*}, \quad z^*=-h^*.
\end{equation}
The bottom motion appears in the right-hand side of Eq. (\ref{kinsolb}). 

The Laplace equation (\ref{lapl}) together with the boundary
conditions (\ref{kinfreesurf}), (\ref{dynfreesurf}) and (\ref{kinsolb}) determine the
boundary-value problem for the velocity potential $\phi^*$ within
the linear theory. In order to solve the equations for a prescribed bottom motion, one
can use Fourier and Laplace transforms (this approach is followed here) or Green's functions. 
Three scenarios are considered for the bottom motion \citep{Dutykh2006a}: 
the passive generation in which the deformation of the sea bottom is simply translated to the free surface
(one is then solving an initial value problem) and two dynamical
processes $\zeta^*(x^*,y^*,t^*)=T(t^*)\zeta^*_{OK}(x^*,y^*)$, where $\zeta^*_{OK}(x^*,y^*)$ is given
by Okada's solution. The two choices for $T$ are
the instantaneous deformation with $T(t^*)=T_i(t^*) = \mathcal{H}(t^*)$, where $\mathcal{H}(t^*)$ denotes
the Heaviside step function, and the exponential law used above in Section 3:
$$    T(t^*) = T_e(t^*) = \left\{%
    \begin{array}{ll}
        0, & t^*<0, \\
        1 - e^{-\alpha^* t^*}, & t^*\geq 0, 
    \end{array}%
    \right. \quad \mbox{with} \;\; \alpha^* > 0.
    $$

Let $\widehat\zeta^*_{OK}$ be the Fourier transform of $\zeta^*_{OK}$,
$\omega^2=g|\k^*|\tanh(|\k^*|h^*)$ the dispersion relation and $\x^*=(x^*,y^*)$.
We provide the general integral solution for the free surface elevation in the three cases, 
\begin{equation}\label{genintsol1}
  \eta_o^*(\x^*,t^*) = \frac{1}{(2\pi)^2}\int\!\!\!\int\limits_{\!\!\!\!\!\R^2}
  {\widehat\zeta^*_{OK} e^{i\k^*\cdot\x^*}} \cos(\omega t^*) \;d\k^*, \quad \mbox{(passive)}
\end{equation}
\begin{equation}\label{genintsol2}
  \eta_i^*(\x^*,t^*) = \frac{1}{(2\pi)^2}\int\!\!\!\int\limits_{\!\!\!\!\!\R^2}
  \frac{\widehat\zeta^*_{OK} e^{i\k^*\cdot\x^*}}{\cosh(|\k^*|h^*)} \cos(\omega t^*) \;d\k^*, \quad \mbox{(instantaneous)}
\end{equation}
\begin{equation}\label{genintsol3}
  \eta_e^*(\x^*,t^*) = \frac{-\alpha^{*2}}{(2\pi)^2}\int\!\!\!\int\limits_{\!\!\!\!\!\R^2}
  \frac{\widehat\zeta^*_{OK} e^{i\k^*\cdot\x^*}}{\cosh(|\k^*|h^*)} \frac{
e^{-\alpha^*t^*}-\cos(\omega t^*)-\omega/\alpha^*\sin(\omega t^*)}{\alpha^{*2}+\omega^2} \;d\k^*,
\end{equation}
and the velocity potential in the passive case (the expressions for the other two active cases are a bit cumbersome)
\citep{Dutykh2006a},
\begin{equation}
  \phi_o^*(\x^*,t^*) = 
  \frac{1}{4\pi^2}\int\!\!\!\int\limits_{\!\!\!\!\!\R^2}
  \frac{-g}{\omega} \widehat\zeta^*_{OK} e^{i\k^*\cdot\x^*}(\cosh|\k^*|z^*+\tanh|\k^*|h^*\sinh|\k^*|z^*)\; d\k^*.
\end{equation}
Then one can easily compute the kinetic and potential energies:
\begin{equation} \label{ekep}
E^*_K = \frac{1}{2}\rho^*\int\!\!\!\int\limits_{\!\!\!\!\!\R^2} \int_{-h^*}^{\eta^*} |\nabla\phi^*|^2 \;d\x^*dz^*, \quad 
E^*_P = \frac{1}{2}\rho^* g\int\!\!\!\int\limits_{\!\!\!\!\!\R^2} \eta^{*2}\;d\x^*.
\end{equation}

\begin{table}
     \begin{center}
       \begin{tabular}{c|l}
         \hline\hline
         \textit{Parameter} & \textit{Value} \\
         \hline
         Dip angle, $\delta$ & $13^\circ$ \\
         \hline
         Slip angle, $\theta$ & $90^\circ$ \\
         \hline
          Fault length, $L^*$ & $150$ km \\
         \hline
         Fault width, $W^*$ & $50$ km \\
         \hline
         Fault depth & $35$ km \\
         \hline
         Slip along the fault & $15$ m \\
         \hline
         Poisson ratio & $0.27$ \\
         \hline
         Young modulus & $9.5\times 10^{9}$ Pa \\
         \hline
         Acceleration due to gravity, $g$ & $9.81$ m/$s^2$ \\
         \hline
         Water depth, $h^*$ & $4$ km \\
         \hline
         Characteristic rise time, $t^*_0$ & $50$ s \\
         \hline
         $\alpha^* = \log(3)/t^*_0$ & $0.0220$ $s^{-1}$ \\
         \hline\hline
       \end{tabular}
       \caption{Values of physical parameters used for the Cauchy-Poisson analysis of
       tsunami generation.}
       \label{tab:CauchyPoisson}
     \end{center}
\end{table}

\begin{figure}
	\centering
		\subfigure[Kinetic energy]%
		{\includegraphics[width=0.49\textwidth]{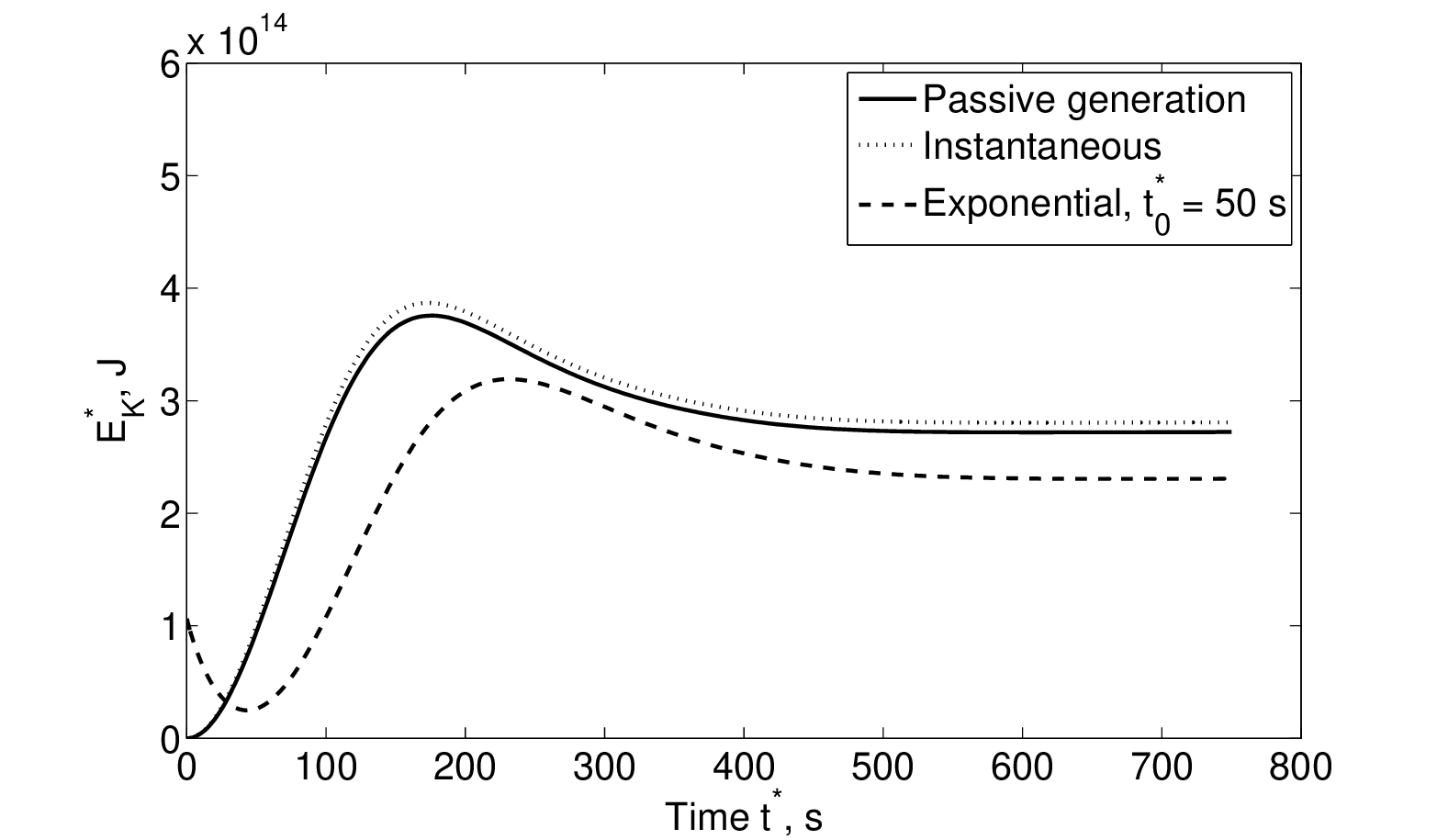}}
		\subfigure[Potential energy]%
		{\includegraphics[width=0.49\textwidth]{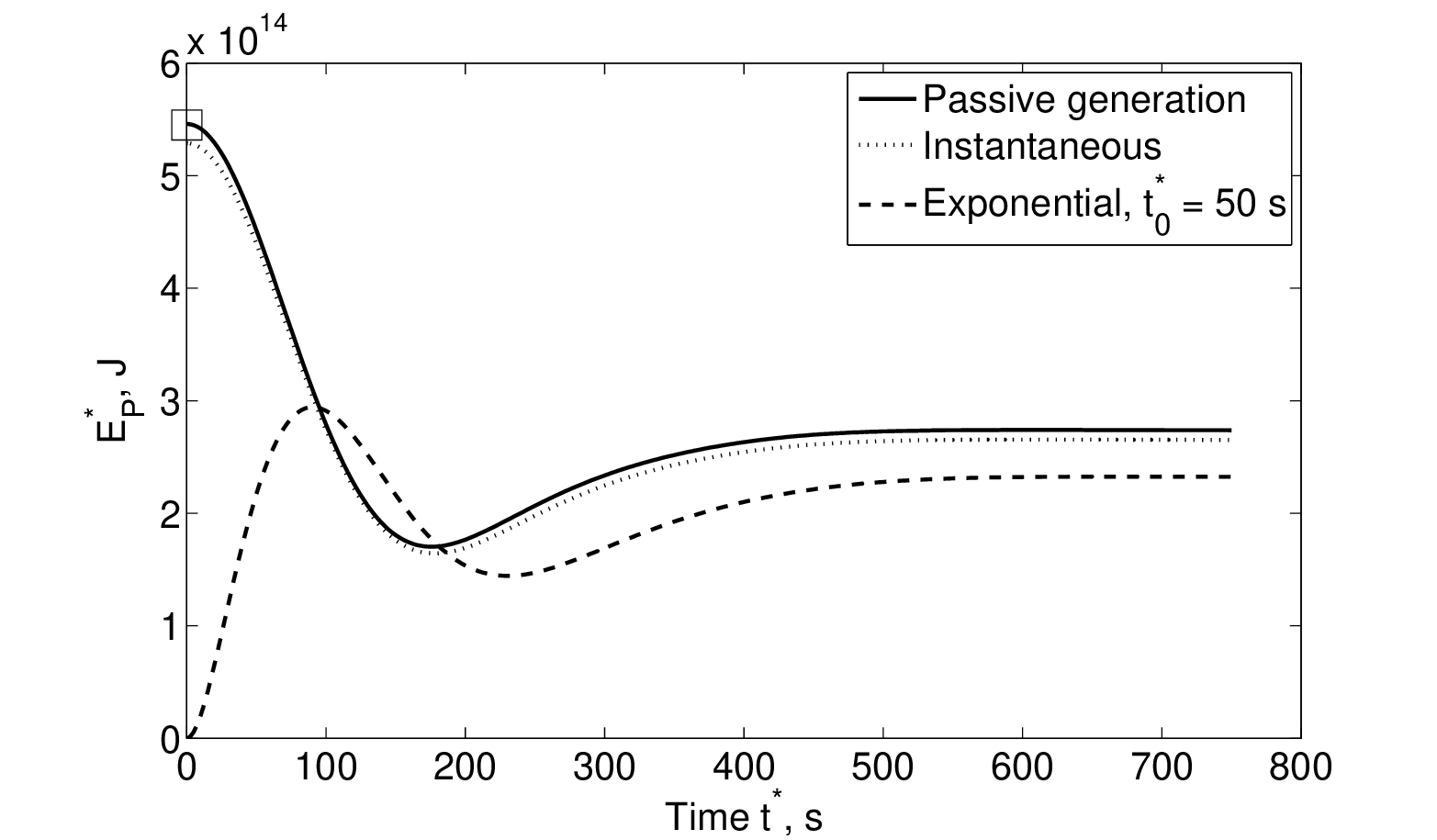}}
		\caption{Time evolution of kinetic and potential energies (\ref{ekep}) for three mechanisms of tsunami generation: passive generation,
instantaneous bottom motion and exponential bottom motion. The square in plot (b) indicates the estimate (\ref{energyOKAL}).}
	\label{CP1}
\end{figure}

\begin{figure}
	\centering
		\subfigure[Total energy]%
		{\includegraphics[width=0.49\textwidth]{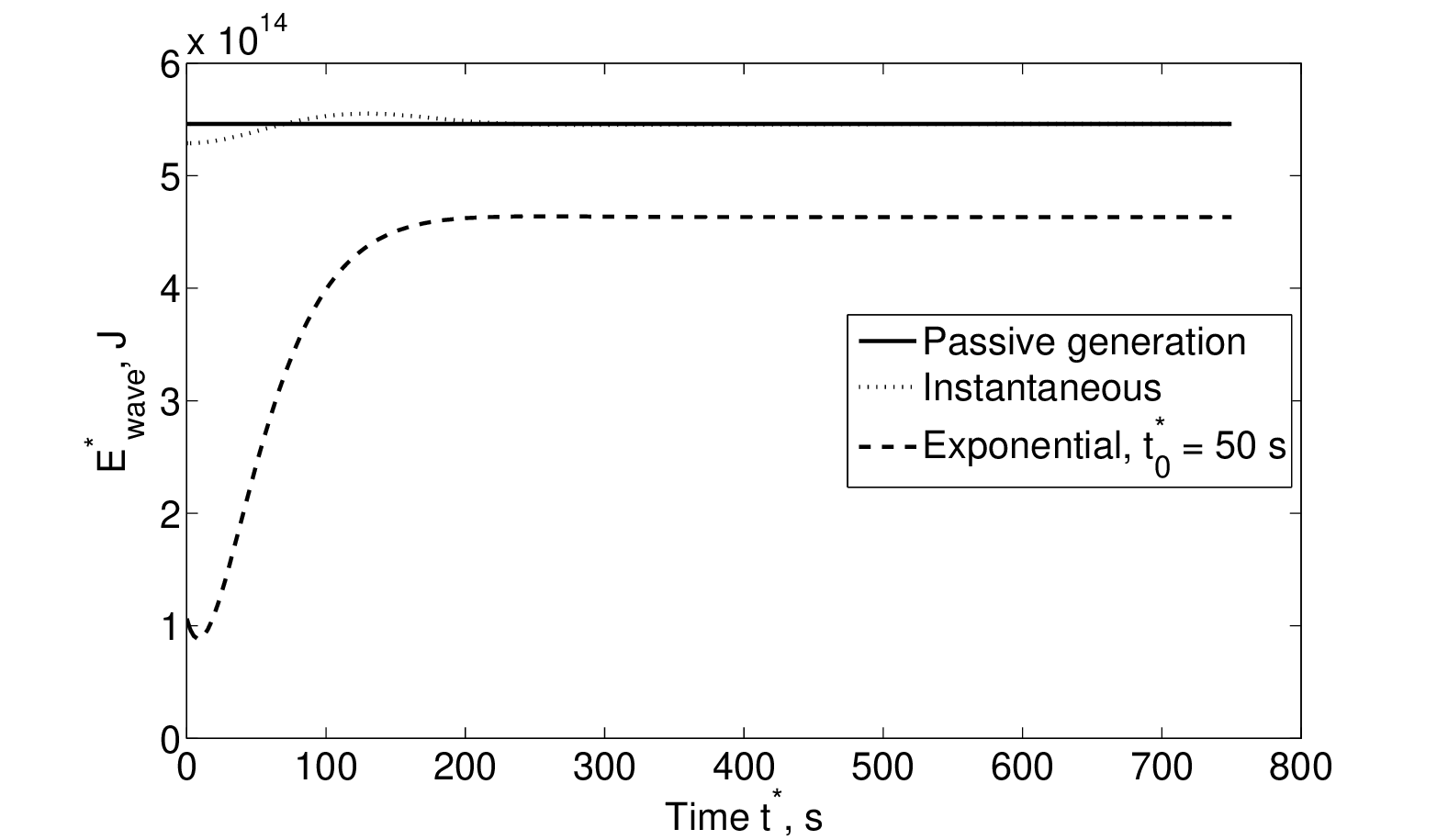}}
		\subfigure[Trajectory in energy space]%
		{\includegraphics[width=0.49\textwidth]{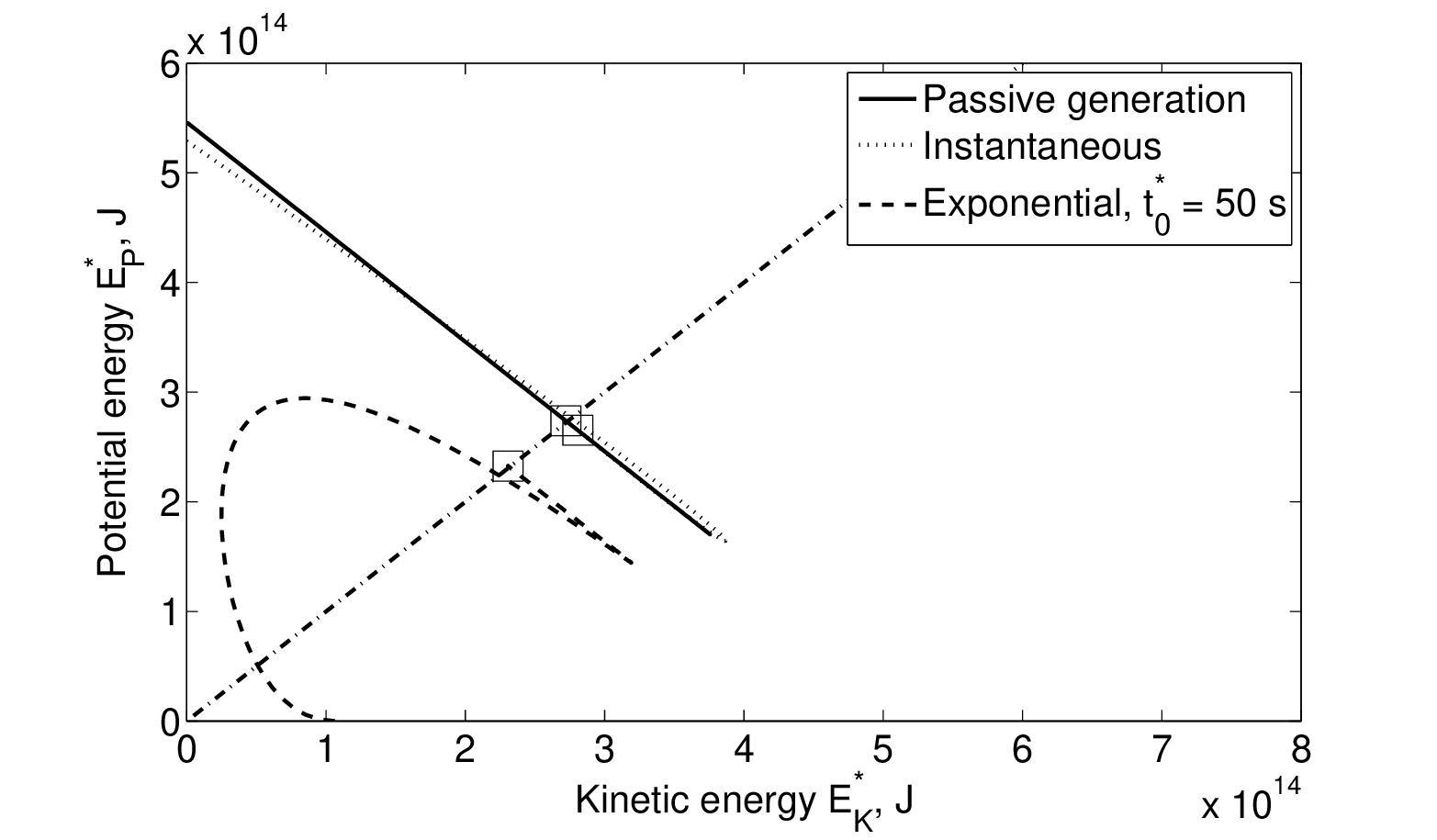}}
		\caption{Same as figure \ref{CP1}. Time evolution of total energy $E_{\rm wave}^* = E^*_K + E^*_P$ (a) and 
trajectory in energy space (b). The straight dashed-dotted line represents equipartition of kinetic and potential energies. 
The squares represent the last computed points.}
	\label{CP2}
\end{figure}

Results are shown in figures \ref{CP1} and \ref{CP2}. Even though there are differences during the first few seconds,
the three mecanisms lead to the same almost exact equipartition between kinetic and potential energy once the dipolar 
waves start to propagate. The simplest estimate proposed for the energy of tsunamis generated by a dislocation source is that given
by \cite{OS03}. They compute the increase in potential energy of the ocean by displacing a volume of water $S \times
\delta h^*$ from the bottom
to the surface of the ocean. This also represents the work of the pressure forces displacing the ocean bottom. Then they
explain that the center of mass of the displaced water, initially at height $\delta h^*/2$ above the ocean floor, is 
transferred to the ocean surface, so that the change in potential energy is not as much. The difference between the two
is the energy available to the tsunami wave:
\begin{equation}
\int\!\!\!\int E_{\rm wave}^* d\x^* = \frac{1}{2}\rho^* g S (\delta h^*)^2. 
\end{equation}
Incidentally, this expression does not depend on the sign of $\delta h^*$ and is also valid for a sudden subsidence
of a section of the ocean floor. It can be extended to a more realistic sea floor deformation, such as the one
used in this paper, $\zeta^*_{OK}(x^*,y^*)$:
\begin{equation} \label{energyOKAL}
\int\!\!\!\int E_{\rm wave}^* d\x^* = \frac{1}{2}\rho^* g \int\!\!\!\int (\zeta^*_{OK})^2 d\x^*. 
\end{equation}
This quantity corresponds to the square in figure \ref{CP1}(b).

\section{Concluding remarks}\label{sec:concl}

In this article we provided a formal derivation of the total energy equation in the framework
of the nonlinear SWE, both for dispersive and non-dispersive waves. We also made an attempt to better 
understand the energy transfer from a moving bottom to the water above.
The importance of this topic is clear, given the serious hazard that tsunamis represent
for coastal regions. 

Tsunami energy can be studied at several levels. A simple formula was given by \cite{OS03}.
In the present paper, we extended it to more spatially realistic sea floor deformations. But this formula does not involve
any dynamics, which can play an important role in tsunami generation. A somewhat counterintuitive
consequence of this simple estimate is that the energy carried by the full tsunami wave is practically 
independent of depth \citep{OS04}. One would think that the energy
involved in lifting a 4 km column of water above a rupture is
different from lifting only 100 m of water (whether instantaneously or in a
few seconds). But no difference can be found in wave height.
The assumption of incompressibility may in fact no longer be valid, especially for a very deep ocean.

The emphasis of this paper has been on tsunami generation. At the runup or inundation stage, energy is also quite
important. Potential energy is being transferred into kinetic energy and the study of these exchanges is left
for future work. Our approach can also be used to analyze the structure of the wave field
in caustics more accurately than with ray theory \citep{Berry}. 
The present paper can be considered as a first step towards a better understanding of tsunami wave energy,
in order to provide scales for tsunami magnitudes for example.
More profound mathematical and physical analysis is needed. 

Another extension is the study of other mechanisms. For example \cite{Okal03} showed that the combination of a lesser absolute level of excitation and a more pronounced
shift of the spectral energy towards higher frequencies characterized by strong dispersion makes landslide
sources significantly deficient far-field tsunami generators, as compared to classical dislocations. 


The authors would like to thank Jean-Michel Ghidaglia
who suggested the idea of considering Boussinesq equations with energy. We are also grateful to him for numerous 
fruitful discussions on numerical
and fluid mechanics topics. The second author acknowledges the support from the EU project TRANSFER (Tsunami Risk ANd Strategies For the European Region) of the sixth Framework Programme under contract no. 037058.

\appendix{Derivation of dispersive shallow-water equations with variable bathymetry}

We derive the Boussinesq equations following the method
used, for example, by \cite{Yoon1989, Villeneuve1993, Nwogu1993}. 
A moving bathymetry was considered in the framework of the Boussinesq equations by \cite{Villeneuve1993}, but they dealt with a 2D problem leading to a 1D system of equations. 

We need to know the depth dependence of the horizontal velocity $\u_{\perp}$ in order to reduce the problem to a 2D one. 
We expand $\u_{\perp}$ in a Taylor series in the vertical coordinate $z$ about the seabed $z=-h$:
\begin{equation}\label{eq:Tayloru}
  \u(x,y,z,t) = \left.\u\right|_{z=-h} + (z+h)\left.\u_z\right|_{z=-h} + 
  \frac{(z+h)^2}{2!}\left.\u_{zz}\right|_{z=-h} + \ldots .
\end{equation}
From now on, the horizontal velocity at the bottom is denoted by
$$
  \u_b := \u_{\perp}(x,y,-h,t).
$$

If the flow is assumed to be irrotational, there is an additional relation which 
closes the system:
\begin{equation}\label{eq:irrotational}
  (\u_{\perp})_z = \nabla w.
\end{equation}
Substituting (\ref{eq:verticalcelerityint}) into (\ref{eq:irrotational}) clearly shows that
$(\u_{\perp})_z(x,y,-h,t) = O(\m)$.
Substituting (\ref{eq:Tayloru}) into (\ref{eq:verticalcelerityint}) and integrating then yields
\begin{equation}\label{eq:vertcelerity}
  w = \m \zeta_t - \m\nabla\cdot\bigl((z+h)\u_b\bigr) + O(\mm).
\end{equation}
The vertical velocity varies 
linearly with respect to $z$ over the depth at leading order $O(\m)$.

The horizontal velocities can be found by integrating the irrotationality condition
(\ref{eq:irrotational}) from $-h$ to $z$:
\begin{equation}\label{eq:irrotint}
  \u_{\perp} = \u_b + \int\limits_{-h}^z\nabla w\; dz.
\end{equation}
Substituting (\ref{eq:vertcelerity}) for $w$ and integrating gives
\begin{multline}\label{eq:horizontcel}
  \u_{\perp} = \u_b + \m(z+h)\left[
  \nabla \zeta_t - \frac{1}{2}(z-h)\nabla(\nabla\cdot\u_b) -
  \nabla\bigl(\nabla\cdot(h\u_b)\bigr)\right] + O(\mu^4).
\end{multline}
We see that the horizontal velocities vary quadratically with respect to $z$ over the depth at leading 
order $O(\m)$.

Following \cite{Ursell1953}, we introduce a number
which measures the relative importance of nonlinear and dispersive effects in long waves:
\begin{equation*}
  S := {\eps}/{\m}.
\end{equation*}
In order to simplify the computations we now assume that $S= O(1)$.
Dispersive terms can be neglected if $S\gg1$.

An expression for the pressure can be obtained 
by substituting (\ref{eq:horizontcel}) and (\ref{eq:vertcelerity}) into (\ref{eq:pressatz}), 
integrating and retaining leading order terms:
\begin{equation}\label{eq:pressfinal}
  \pi = \eta + \frac{\m}{2}z^2\nabla\cdot\u_{bt} + 
  \m z\bigl(\nabla\cdot(h\u_{bt})-\zeta_{tt}\bigr) + O(\mm).
\end{equation}

The equation for the free-surface evolution is derived by substituting (\ref{eq:horizontcel}) into the
depth-integrated continuity equation (\ref{eq:contint}) and integrating:
\begin{equation}\label{eq:freesurfb}
  \eta_t + \nabla\cdot\bigl((h+\eps\eta)\u_b\bigr) -\m
  \nabla\cdot\Bigl[\frac{h^2}{2}\nabla\bigl(\nabla\cdot(h\u_b)-\zeta_t\bigr)
  -\frac{h^3}{3}\nabla\bigl(\nabla\cdot\u_b\bigr)\Bigr] - \zeta_t = 0.
\end{equation}

The equation for the evolution of the horizontal velocity is obtained by substituting (\ref{eq:horizontcel}) and
(\ref{eq:pressfinal}) into (\ref{eq:xmomentum}) and (\ref{eq:ymomentum}):
\begin{multline*}
  \u_{bt} + \eps (\u_b\cdot\nabla)\u_b + \nabla\eta + \m\Bigl[
  \frac{h^2}{2}\nabla(\nabla\cdot\u_{b}) - h\nabla\bigl(\nabla\cdot(h\u_{b})-\zeta_{t}\bigr)\Bigr]_t = 0.
\end{multline*}
Using the irrotationality condition this equation can be rewritten as
\begin{multline}\label{eq:horvelocitybot}
  \u_{bt} + \frac{\eps}{2}\nabla\abs{\u_b}^2 + \nabla\eta + \m\Bigl[
  \frac{h^2}{2}\nabla(\nabla\cdot\u_{b}) - h\nabla\bigl(\nabla\cdot(h\u_{b})-\zeta_{t}\bigr)\Bigr]_t = 0.
\end{multline}

Finally we write the energy equation by
substituting (\ref{eq:horizontcel}) and (\ref{eq:pressfinal}) into (\ref{eq:energyint}):
\begin{equation}\label{eq:energybot}
  \pd{}{t}\int\limits_{-h}^{\eps\eta} e\; dz +
  \eps\nabla\cdot\left[\left(\int\limits_{-h}^{\eps\eta}e\;dz + \frac{1}{2}(h+\eps\eta)^2\right)\u_b\right]
  - \eps (h+\eps\eta) \zeta_t = 0.
\end{equation}  
The higher-order terms are of order $O(\eps\m)$.

Another possibility for the choice of variables is to introduce the
depth averaged velocity. The corresponding standard Boussinesq-type equations were obtained by \cite{Peregrine1967} in the case
of a fixed seabed. We extend the results to a moving bathymetry and add the energy equation.
The main advantage of the depth averaged velocity consists in the fact that
the continuity equation (or equivalently the equation for the free-surface elevation) is very simple and exact in this variable.

Let us rewrite all equations in terms of the depth-averaged velocity defined by
\begin{equation}\label{eq:defdepthaveraged}
  \bar\u = \frac{1}{h+\eps\eta}\int\limits_{-h}^{\eps\eta} \u_{\perp} \;dz.
\end{equation}

The depth-integrated continuity equation (\ref{eq:contint}) yields immediately
\begin{equation}\label{eq:contdepthaveraged}
  \eta_t + \nabla\cdot\bigl((h+\eps\eta)\bar\u\bigr) - \zeta_t = 0.
\end{equation}

In order to derive equations for the horizontal velocity and the energy we need
a relation between $\u_b$ and $\bar\u$. The desired relation is deduced
directly from the definition (\ref{eq:defdepthaveraged}) by substituting 
(\ref{eq:horizontcel}) in it:
\begin{equation*}
  \bar\u = \u_b - \frac{\m}{2}h\nabla\bigl(\nabla\cdot(h\u_b)-\zeta_t\bigr) +
  \frac{\m}{3}h^2\nabla(\nabla\cdot\u_b) + O(\eps^2 + \eps\m + \mm).
\end{equation*}
Inverting the last equation yields
\begin{equation}\label{eq:relationubarub}
  \u_b = \bar\u + \m \Bigl[\frac{h}{2}\nabla\bigl(\nabla\cdot(h\u_b)-\zeta_t\bigr) -
  \frac{h^2}{3}\nabla(\nabla\cdot\u_b)\Bigr] + O(\eps^2 + \eps\m + \mm).
\end{equation}

Substituting the relation (\ref{eq:relationubarub}) into equation 
(\ref{eq:horvelocitybot}) gives the standard Boussinesq equations for a moving bottom:
\begin{equation}\label{eq:horveldepthaveraged}
  \bar\u_t + \frac{\eps}{2}\nabla\abs{\bar\u}^2 + \nabla\eta +
  \m\Bigl[\frac{h^2}{6}\nabla(\nabla\cdot\bar\u) - \frac{h}{2}
  \nabla(\nabla\cdot(h\bar\u)-\zeta_{t})\Bigr]_t = 0.
\end{equation}

The energy equation is obtained by substituting the relation (\ref{eq:relationubarub}) into
equation (\ref{eq:energybot}):
\begin{equation}\label{eq:energyubar}
  \pd{}{t}\int\limits_{-h}^{\eps\eta} e\; dz +
  \eps\nabla\cdot\left[\left(\int\limits_{-h}^{\eps\eta}e\;dz + \frac{1}{2}(h+\eps\eta)^2\right)\bar\u\right]
  - \eps (h+\eps\eta) \zeta_t = 0.
\end{equation}


%
%


Since the energy equation is redundant for incompressible flows, the linear dispersion relation is unaffected by
the inclusion of the energy equation. As is well-known, it can be improved by defining the horizontal velocity at an arbitrary level. 


\bibliography{energy}

\begin{thebibliography}{34}
\providecommand{\natexlab}[1]{#1}
\providecommand{\url}[1]{\texttt{#1}}
\expandafter\ifx\csname urlstyle\endcsname\relax
  \providecommand{\doi}[1]{doi: #1}\else
  \providecommand{\doi}{doi: \begingroup \urlstyle{rm}\Url}\fi

\bibitem[Ben-Menahem and Rosenman(1972)]{Ben-M}
A.~Ben-Menahem and M.~Rosenman.
\newblock Amplitude patterns of tsunami waves from submarine earthquakes.
\newblock \emph{J. Geophys. Res.}, 77:\penalty0 3097--3128, 1972.

\bibitem[Berry(2007)]{Berry}
M.V. Berry.
\newblock Focused tsunami waves.
\newblock \emph{Proc. R. Soc. A}, 463:\penalty0 3055--3071, 2007.

\bibitem[Dao and Tkalich(2007)]{Dao2007}
M.H. Dao and P.~Tkalich.
\newblock Tsunami propagation modelling - a sensivity study.
\newblock \emph{Nat. Hazards Earth Syst. Sci.}, 7:\penalty0 741--754, 2007.

\bibitem[Dotsenko and Korobkova(1997)]{Dotsenko1997}
S.F. Dotsenko and T.Yu. Korobkova.
\newblock The effect of frequency dispersion on plane waves generated as a
  result of bed motions.
\newblock \emph{Phys. Oceanogr.}, 8(3):\penalty0 143--154, 1997.

\bibitem[Dutykh and Dias(2007)]{Dutykh2006}
D.~Dutykh and F.~Dias.
\newblock Water waves generated by a moving bottom.
\newblock In Anjan Kundu, editor, \emph{Tsunami and Nonlinear Waves}, pages
  63--94. Springer Verlag (Geo Sc.), 2007.

\bibitem[Dutykh et~al.(2006)Dutykh, Dias, and Kervella]{Dutykh2006a}
D.~Dutykh, F.~Dias, and Y.~Kervella.
\newblock Linear theory of wave generation by a moving bottom.
\newblock \emph{C. R. Acad. Sci. Paris, Ser. I}, 343:\penalty0 499--504, 2006.

\bibitem[Duytkh et~al.(2008)Duytkh, Poncet, and Dias]{VOLNA}
D.~Duytkh, R.~Poncet, and F.~Dias.
\newblock Complete numerical modelling of tsunami waves: generation,
  propagation and inundation.
\newblock \emph{in preparation}, 2008.

\bibitem[Eskilsson and Sherwin(2005)]{Eskilsson}
C.~Eskilsson and S.J. Sherwin.
\newblock Discontinuous {G}alerkin spectral/hp element modelling of dispersive
  shallow water systems.
\newblock \emph{J. Sci. Comp.}, 22-23:\penalty0 269--288, 2005.

\bibitem[Gisler(2008)]{Gisler2008}
G.R. Gisler.
\newblock Tsunami simulations.
\newblock \emph{Annu. Rev. Fluid Mech.}, 40:\penalty0 71--90, 2008.

\bibitem[Hammack(1973)]{Hammack1973}
J.~Hammack.
\newblock A note on tsunamis: their generation and propagation in an ocean of
  uniform depth.
\newblock \emph{Journal of Fluid Mechanics}, 60:\penalty0 769--799, 1973.

\bibitem[Johnson(1997)]{Johnson1997}
R.~S. Johnson.
\newblock \emph{A modern introduction to the mathematical theory of water
  waves}.
\newblock Cambridge University Press, Cambridge, 1997.

\bibitem[Kajiura(1970)]{Kajiura1970}
K.~Kajiura.
\newblock Tsunami source, energy and the directivity of wave radiation.
\newblock \emph{Bulletin of the Earthquake Research Institute}, 48:\penalty0
  835--869, 1970.

\bibitem[K\^anoglu and Synolakis(2006)]{KS2006}
U.~K\^anoglu and C.E. Synolakis.
\newblock Initial value problem solution of nonlinear shallow water-wave
  equations.
\newblock \emph{Phys. Rev. Lett.}, 97:\penalty0 148501, 2006.

\bibitem[Kervella et~al.(2007)Kervella, Dutykh, and Dias]{Kervella2007}
Y.~Kervella, D.~Dutykh, and F.~Dias.
\newblock Comparison between three-dimensional linear and nonlinear tsunami
  generation models.
\newblock \emph{Theoretical and Computational Fluid Dynamics}, 21:\penalty0
  245--269, 2007.

\bibitem[Kowalik et~al.(2007)Kowalik, Knight, Logan, and Whitmore]{Kowalik2007}
Z.~Kowalik, W.~Knight, T.~Logan, and P.~Whitmore.
\newblock The tsunami of 26 {D}ecember, 2004: {N}umerical modeling and energy
  considerations.
\newblock \emph{Pure and Applied Geophysics}, 164:\penalty0 379--393, 2007.

\bibitem[Kulikov et~al.(2005)Kulikov, Medvedev, and Lappo]{Kulikov2005}
E.~Kulikov, P.~Medvedev, and S.~Lappo.
\newblock Dispersion of the {S}umatra tsunami waves in the {I}ndian {O}cean
  detected by satellite altimetry.
\newblock \emph{Trans. (Doklady) Russian Acad. Sci., Oceanology}, 401:\penalty0
  537--542, 2005.

\bibitem[Kulikov et~al.(1996)Kulikov, Rabinovich, Thomson, and
  Bornhold]{Kulikov1996}
E.A. Kulikov, A.B. Rabinovich, R.E. Thomson, and B.D. Bornhold.
\newblock The landslide tsunami of {N}ovember 3, 1994, {S}kagway {H}arbor,
  {A}laska.
\newblock \emph{J. Geophys. Res.}, 101(C3):\penalty0 6609--6615, 1996.

\bibitem[Nwogu(1993)]{Nwogu1993}
O.~Nwogu.
\newblock Alternative form of {B}oussinesq equations for nearshore wave
  propagation.
\newblock \emph{J. Waterways Port Coastal Ocean Engng, ASCE}, 119:\penalty0
  618--638, 1993.

\bibitem[Okada(1985)]{Okada85}
Y.~Okada.
\newblock Surface deformation due to shear and tensile faults in a half-space.
\newblock \emph{Bull. Seism. Soc. Am.}, 75:\penalty0 1135--1154, 1985.

\bibitem[Okal(2003)]{Okal03}
E.~A. Okal.
\newblock Normal mode energetics for far-field tsunamis generated by
  dislocations and landslides.
\newblock \emph{Pure Appl. Geophys.}, 160:\penalty0 2189--2221, 2003.

\bibitem[Okal and Synolakis(2003)]{OS03}
E.~A. Okal and C.~E. Synolakis.
\newblock A theoretical comparison of tsunamis from dislocations and
  landslides.
\newblock \emph{Pure Appl. Geophys.}, 160:\penalty0 2177--2188, 2003.

\bibitem[Okal and Synolakis(2004)]{OS04}
E.~A. Okal and C.~E. Synolakis.
\newblock Source discriminants for near-field tsunamis.
\newblock \emph{Geophys. J. Int.}, 158:\penalty0 899--912, 2004.

\bibitem[Okal et~al.(2007)Okal, Talandier, and Reymond]{OTR2007}
E.~A. Okal, J.~Talandier, and D.~Reymond.
\newblock Quantification of hydrophone records of the 2004 {S}umatra tsunami.
\newblock \emph{Pure Appl. Geophys.}, 164:\penalty0 309--323, 2007.

\bibitem[Peregrine(1967)]{Peregrine1967}
D.~H. Peregrine.
\newblock Long waves on a beach.
\newblock \emph{J. Fluid Mech.}, 27:\penalty0 815--827, 1967.

\bibitem[Ritsema et~al.(1995)Ritsema, Ward, and Gonz\'alez]{Ritsema1995}
J.~Ritsema, S.~N. Ward, and F.~I. Gonz\'alez.
\newblock Inversion of deep-ocean tsunami records for 1987 to 1988 {G}ulf of
  {A}laska earthquake parameters.
\newblock \emph{Bulletin of the Seismological Society of America}, 85:\penalty0
  747--754, 1995.

\bibitem[Stoker(1958)]{Stoker1958}
J.J. Stoker.
\newblock \emph{Water waves, the mathematical theory with applications}.
\newblock Wiley, 1958.

\bibitem[Synolakis and Bernard(2006)]{Syno2006}
C.E. Synolakis and E.N. Bernard.
\newblock Tsunami science before and beyond {B}oxing {D}ay 2004.
\newblock \emph{Phil. Trans. R. Soc. A}, 364:\penalty0 2231--2265, 2006.

\bibitem[Tinti and Bortolucci(2000)]{Tinti2000}
S.~Tinti and E.~Bortolucci.
\newblock Energy of water waves induced by submarine landslides.
\newblock \emph{Pure Appl. Geophys.}, 157:\penalty0 281--318, 2000.

\bibitem[Titov et~al.(2005)Titov, Rabinovich, Mofjeld, Thomson, and
  Gonz\'alez]{Titov}
V.V. Titov, A.B. Rabinovich, H.O. Mofjeld, R.E. Thomson, and F.I. Gonz\'alez.
\newblock The global reach of the 26 {D}ecember 2004 {S}umatra tsunami.
\newblock \emph{Science}, 309:\penalty0 2045--2048, 2005.

\bibitem[Ursell(1953)]{Ursell1953}
F.~Ursell.
\newblock The long-wave paradox in the theory of gravity waves.
\newblock \emph{Proc. Camb. Phil. Soc.}, 49:\penalty0 685--694, 1953.

\bibitem[Velichko et~al.(2002)Velichko, Dotsenko, and Potetyunko]{Velichko2002}
A.S. Velichko, S.F. Dotsenko, and \'E.N. Potetyunko.
\newblock Amplitude-energy characteristics of tsunami waves for various types
  of seismic sources generating them.
\newblock \emph{Phys. Oceanogr.}, 12(6):\penalty0 308--322, 2002.

\bibitem[Villeneuve and Savage(1993)]{Villeneuve1993}
M.~Villeneuve and S.B. Savage.
\newblock Nonlinear, dispersive, shallow-water waves developed by a moving bed.
\newblock \emph{Journal of Hydraulic Research}, 31\penalty0 (2):\penalty0
  249--266, 1993.

\bibitem[Ward(1980)]{Ward80}
S.N. Ward.
\newblock Relationship of tsunami generation and an earthquake source.
\newblock \emph{J. Phys. Earth}, 28:\penalty0 441--474, 1980.

\bibitem[Yoon and Liu(1989)]{Yoon1989}
S.B. Yoon and P.L.-F. Liu.
\newblock Interactions of currents and weakly nonlinear water waves in shallow
  water.
\newblock \emph{J. Fluid Mech.}, 205:\penalty0 397--419, 1989.

\end{thebibliography}
\bibliographystyle{plainnat}
\end{document}